\documentclass[11pt]{article}
\usepackage[T1]{fontenc}
\usepackage{lmodern}
\usepackage[margin=1in]{geometry}
\usepackage{amsmath,amssymb,amsthm,mathtools}
\usepackage{microtype,booktabs,enumitem,xcolor}
\usepackage{needspace}
\usepackage{float}
\floatstyle{ruled}
\newfloat{algorithm}{tbp}{loa}
\floatname{algorithm}{Algorithm}
\usepackage[colorlinks=true,linkcolor=blue!45!black,citecolor=blue!45!black,urlcolor=blue!45!black]{hyperref}
\usepackage[nameinlink,noabbrev]{cleveref}
\numberwithin{equation}{section}
\newtheorem{theorem}{Theorem}[section]
\newtheorem{lemma}[theorem]{Lemma}
\newtheorem{proposition}[theorem]{Proposition}
\newtheorem{corollary}[theorem]{Corollary}
\theoremstyle{definition}
\newtheorem{definition}[theorem]{Definition}
\theoremstyle{remark}
\newtheorem{remark}[theorem]{Remark}
\DeclareMathOperator{\diag}{diag}
\DeclareMathOperator{\supp}{supp}
\DeclareMathOperator{\tr}{tr}

\newcommand{\R}{\mathbb R}
\newcommand{\1}{\mathbf1}
\newcommand{\softO}{\widetilde O}
\newcommand{\Call}[1]{\textsc{#1}}
\crefname{algorithm}{Algorithm}{Algorithms}
\Crefname{algorithm}{Algorithm}{Algorithms}
\setlist{topsep=4pt,itemsep=3pt,parsep=0pt}
\allowdisplaybreaks[2]

\hypersetup{pdftitle={Directed Singular-Value Sparsifiers with O(n log n / epsilon squared) Arcs},pdfauthor={Emile Anand}}
\title{Directed Singular-Value Sparsifiers with $O(n\log n/\varepsilon^2)$ Arcs}
\author{Emile Anand\\ Georgia Institute of Technology}
\date{}
\begin{document}
\maketitle
\begin{abstract}
Given an unweighted directed graph, a sparsifier is a nonnegative reweighting of original arcs that preserves every in-degree and out-degree exactly. In this work, we study singular-value sparsifiers, which are generalizations of spectral sparsifiers. Singular-value sparsifiers find a number of applications in theoretical computer science, such as maintaining directed cuts and flows, speeding up directed Laplacian solvers, and sparsifying dense weight matrices in neural networks. 
Strong singular-value (SV) approximation supports sparse representations
of directed random walks with guarantees that remain stable under
products, and controls bilinear error in the left and right
Schur-complement metrics. We prove that every unweighted directed graph on $n$ vertices
admits an $\varepsilon$-SV sparsifier with
$O(n\log(2n)/\varepsilon^2)$ arcs, improving the
$O(n\log^5 n/\varepsilon^2)$ support bound of Lau, Wang, and Zhou \cite{lwz}.
The sparsifier is a nonnegative reweighting of original arcs and
preserves every in-degree and out-degree exactly.

We also give a randomized construction with
$\softO(m+n)$ real-arithmetic running time, where
$m$ is the input support size. The output has the same support bound
and rational weights of polynomial bit length. The key estimate bounds
the variance after we project reweightings onto both sets of degree
constraints. Each resistance piece contributes at most its own
Schur-complement defect, so we can sum the variances without a loss
from overlapping vertices. In dense phases, expander pruning also
controls the covariance of the matrix entries. A Gaussian matrix
bound and a rotation argument then show that projection onto the
degree-constrained cube has the required spectral discrepancy.
We compute this projection with a sparse Jacobi iteration whose cost
grows only logarithmically in the required precision. In the sparser
phases we instead use constrained spectral projection, which is
already cheap there. Each phase ends by rounding weights around
cycles, which deletes arcs while keeping every degree exact.
For positive input weights of ratio $U$,
the existence bound extends to
$O(n\log(2n)/\varepsilon^2+n\log(2U))$.
\end{abstract}

\newpage

\section{Introduction and main results}

Spectral sparsification compresses a graph while preserving the linear
algebra that algorithms use to process it. For an undirected graph,
approximating every Laplacian quadratic form simultaneously preserves
cut values and gives a preconditioner for linear systems. The work of
Spielman and Teng~\cite{st} established this connection.
Spielman and Srivastava~\cite{ss} obtained
$O(n\log n/\varepsilon^2)$ edges by sampling according to effective
resistance, and Batson, Spielman, and Srivastava~\cite{bss} reduced
the support to $O(n/\varepsilon^2)$. These bounds provide a natural
benchmark for directed graphs, where a sparse approximation must also
preserve the information carried by edge orientations.

Random walks show why the directed case is harder. A quadratic form
cannot tell a matrix apart from its transpose, so it does not see which
way the arcs point. On top of that, an error that is small for a single
step of a walk can grow over many steps, so an approximation meant for
long walks has to control how errors build up under products. Strong singular-value (SV) approximation,
introduced by Ahmadinejad, Peebles, Pyne, Sidford, and
Vadhan~\cite{appsv}, provides such control. It raises a basic
sparsification question: how many original arcs suffice to preserve this
stronger guarantee?

We represent a directed graph by its nonnegative adjacency matrix $A$,
with $A_{uv}$ equal to the weight of the arc from $u$ to $v$.
To see what SV approximation asks for, let
$D_L=\diag(A\1)$ and $D_R=\diag(A^\top\1)$, discard zero-degree
coordinates, and write $P=D_L^{-1/2}AD_R^{-1/2}$. For an approximation
$\widetilde A$ with the same degree matrices, define $\widetilde P$
in the same way. The normalized SV inequality asks that every pair
of vectors $x,y$ satisfies
\begin{equation}\label{eq:intro-sv}
 \bigl|x^\top(\widetilde P-P)y\bigr|
 \le \varepsilon
       \sqrt{\|x\|_2^2-\|P^\top x\|_2^2}\,
       \sqrt{\|y\|_2^2-\|Py\|_2^2}.
\end{equation}
The two factors measure how much the forward and reverse operators
contract the test vectors. If a transition preserves a direction
exactly, the allowed error in that direction is zero. For example,
when $P$ is a permutation matrix, \eqref{eq:intro-sv} forces
$\widetilde P=P$. This requirement captures deterministic transitions
without assuming that the walk mixes rapidly.
\Cref{def:sv} gives the equivalent definition for unnormalized
adjacency matrices using two Schur-complement defects.

We prove that $O(n\log(2n)/\varepsilon^2)$ arcs suffice for every
unweighted directed graph, and give a randomized construction that
runs in nearly linear real-arithmetic time, with no polynomial
dependence on $1/\varepsilon$.
The sparsifier preserves every weighted in-degree and out-degree
exactly. The input need not be Eulerian, and the output uses
nonnegative weights on original arcs.

\subsection{Main results}

\begin{theorem}[Unweighted directed inputs]\label{thm:main}
There is an absolute constant $C$ such that for every
$A\in\{0,1\}^{n\times n}$ and $0<\varepsilon<1$, there is a nonnegative
$\varepsilon$-SV approximation $\widetilde A$ to $A$ with
$\supp\widetilde A\subseteq\supp A$, $\widetilde A\1=A\1$, and
$\widetilde A^\top\1=A^\top\1$, whose support size satisfies
\[
 |\supp\widetilde A|
 \le \min\left\{|\supp A|,
           \left\lceil C\frac{n\log(2n)}{\varepsilon^2}\right\rceil\right\}.
\]
We can choose rational output weights with polynomial bit length
in $n$ and $\log(1/\varepsilon)$.
\end{theorem}

\Needspace{9\baselineskip}
\begin{theorem}[Weighted directed inputs]\label{thm:weighted}
For every $0<\varepsilon<1$ and every nonnegative $n\times n$ matrix $A$
whose positive entries have ratio at most $U\ge1$, there is a nonnegative $\varepsilon$-SV
approximation with the same row and column sums, supported on $A$, and
with at most
\[
 \min\left\{|\supp A|,
  \left\lceil C\left(\frac{n\log(2n)}{\varepsilon^2}
                         +n\log(2U)\right)\right\rceil\right\}
\]
nonzero entries. In particular, polynomially bounded weight ratio gives
$O(n\log(2n)/\varepsilon^2)$ support.
\end{theorem}

\begin{theorem}[Randomized construction in real arithmetic]
\label{thm:fast}
Let $A\in\{0,1\}^{n\times n}$ have $m$ nonzero entries, and let
$0<\varepsilon<1$ and $0<\eta<1/2$.
Given the input as an arc list, there is a randomized algorithm that
runs in $\softO(m+n)$ real-arithmetic time and, with probability at
least $1-\eta$, returns a matrix satisfying every conclusion of \cref{thm:main}.
Here $\softO$ hides factors polynomial in
$\log(2n)$, $\log(1/\varepsilon)$, and $\log(1/\eta)$.
\Cref{alg:fast-sv,alg:hybrid-phase} give the algorithm.
\end{theorem}

We use the real-arithmetic model of the optimization algorithms
in~\cite{jrt}, together with scalar comparisons and grid rounding.
We implement matrix operations through sparse products and approximate
Laplacian inverses, and include their costs in the bound of \cref{thm:fast}.
The numerical projection routine has logarithmic precision
dependence. We use a conservative split between dense and sparse
phases, so the hidden powers of $\log n$ are large. A word-RAM bound
would require a separate analysis of the
intermediate computations, even though the final weights have
polynomial bit length. Similar precision questions arise for dynamic
algebraic algorithms~\cite{abgz}.

Lau, Wang, and Zhou~\cite[Theorem~1.4]{lwz} obtained
$O(n\log^5 n/\varepsilon^2)$ support for the same SV guarantee.
\Cref{thm:main,thm:weighted} reduce this to one logarithm for
unweighted inputs and for polynomially bounded weight ratio.
\Cref{thm:fast} makes the unweighted bound constructive in nearly
linear time with polylogarithmic accuracy dependence. Their algorithm is deterministic,
whereas ours is randomized.
In the other direction, \cref{prop:lower} shows that the all-ones
matrix, whose bipartite lift is the complete bipartite graph, has no
$\varepsilon$-SV approximation with fewer than
$\tfrac12\min\{n^2,n/\varepsilon^2\}$ nonzero entries. Thus the support bound leaves a logarithmic gap in the
regime where sparsification is possible.

\subsection{Consequences for queries and random walks}

The guarantee controls the total weight from any vertex set to any
other vertex set. Write
$A(S,T)=\sum_{u\in S,v\in T}A_{uv}$, and let $\1_S,\1_T$ denote the
indicator vectors. With the defects $R_L(A)$ and $R_R(A)$ from
\cref{sec:sv-lift}, \cref{def:sv} gives
\begin{align}
 \bigl|\widetilde A(S,T)-A(S,T)\bigr|
 &=\bigl|\1_S^\top(\widetilde A-A)\1_T\bigr|\notag\\
 &\le\varepsilon
       \sqrt{\1_S^\top R_L(A)\1_S}\,
       \sqrt{\1_T^\top R_R(A)\1_T}.
 \label{eq:intro-sets}
\end{align}
One sparsifier satisfies this bound simultaneously for all $S,T$.
The error scale adapts to the two sets through their defect energies.
We can answer such queries by scanning
$O(n\log(2n)/\varepsilon^2)$ stored arcs after preprocessing.

For an Eulerian input, exact degree preservation also preserves a
stationary distribution of the walk. Discard isolated vertices and
let $D=D_L=D_R$,
$W=D^{-1}A$, $Z=\1^\top D\1$, and $\pi_v=D_{vv}/Z$.
Both $W$ and $\widetilde W=D^{-1}\widetilde A$ have stationary
distribution $\pi$. If each walk starts from $\pi$, then
$\Pr_A[X_0\in S,\ X_1\in T]=A(S,T)/Z$, and the same identity holds
for $\widetilde A$. Hence \eqref{eq:intro-sets} gives
\begin{align*}
 \bigl|\Pr_{\widetilde A}[X_0\in S,\ X_1\in T]
       -\Pr_A[X_0\in S,\ X_1\in T]\bigr|
 &=\frac1Z\bigl|\widetilde A(S,T)-A(S,T)\bigr|\\
 &\le\frac{\varepsilon}{Z}
       \sqrt{\1_S^\top R_L(A)\1_S}\,
       \sqrt{\1_T^\top R_R(A)\1_T}.
\end{align*}
Thus \eqref{eq:intro-sets} directly controls stationary transition
mass between sets.

For $0<\varepsilon\le1/4$, the product theorem of Ahmadinejad
et al.~\cite[Lemma~3.17]{appsv} gives $O(\varepsilon)$ normalized
SV approximation between
$\widetilde P^\ell$ and $P^\ell$ for every $\ell\ge1$, with a constant
independent of $\ell$. For Eulerian inputs,
$P=D^{1/2}WD^{-1/2}$, so these powers describe the same walk in
stationary coordinates. The sparse one-step operator therefore
keeps a guarantee for every walk length. Each application of
$\widetilde W$ to a vector costs
$O(n\log(2n)/\varepsilon^2)$ arithmetic operations. Sparsity is the
simplest structure that makes matrix-vector products cheap.
Anand, van den Brand, and McCarty~\cite{abm} study other structural
parameters of dense matrices with the same effect.

These consequences explain the value of a smaller SV sparsifier.
Fast algorithms that explicitly sparsify long weighted walk powers
also need to control intermediate weights and construct weighted
sparsifiers. Our weighted theorem provides a support bound, while
\cref{thm:fast} provides the stated running time for unweighted inputs.

\subsection{Related work}

The development of directed spectral approximation has followed two
closely connected algorithmic goals: solving directed Laplacian
systems and approximating long random walks.
Cohen, Kelner, Peebles, Peng, Rao, Sidford, and Vladu~\cite{ckp}
introduced a directed approximation notion that supports
sparsification and fast Eulerian Laplacian solvers. Through their
reductions, these solvers give algorithms for stationary
distributions and expected commute times. Cohen, Kelner, Kyng,
Peebles, Peng, Rao, and Sidford~\cite{ckkpprs} later solved general
directed Laplacian systems in nearly linear time through sparse LU
factorizations. Ahmadinejad, Kelner,
Murtagh, Peebles, Sidford, and Vadhan~\cite{akm} introduced
unit-circle approximation in their work on estimating random-walk
probabilities in small space. Their approximation remains valid
under powering, a property that connects spectral methods to
derandomization. The pseudorandomness of random walks on expanders
is a recurring theme there, see for example~\cite{au-sticky,a-acc0}.

SV approximation strengthens these earlier notions.
Ahmadinejad et al.~\cite{appsv} proved a reduction through the
undirected bipartite lift and constructed nearly linear-sized
SV sparsifiers. Their applications include sparse approximations to
long walks and estimates of stationary cut probabilities.
Lau, Wang, and Zhou~\cite{lwz} then used constrained matrix
discrepancy to obtain the $O(n\log^5 n/\varepsilon^2)$ support
bound. Their treatment of expander pieces and exact linear
constraints is a direct predecessor of the present work.

Exact degrees create an additional constraint on the construction.
Independent edge sampling does not generally preserve them.
Chu, Gao, Peng, Sachdeva, Sawlani, and Wang~\cite{cgp} developed
short cycle decompositions to obtain degree-preserving sparsifiers
and improved directed sparsifiers. Alternating updates around a
cycle preserve the degree equations while deleting edges.
Jambulapati, Reis, and Tian~\cite{jrt} brought discrepancy methods
to nearly linear-time sparsification and obtained linear-sized
degree-preserving sparsifiers for undirected graphs. We use their
constrained optimization framework in phases where its inverse-radius
cost is polylogarithmic. For denser phases, we compute a Gaussian
projection onto the degree-constrained cube by a sparse Jacobi
iteration and exact tree correction.

For the Eulerian approximation notion, Sachdeva, Thudi, and
Zhao~\cite{stz} obtained sparser constructions through discrepancy.
Jambulapati, Sachdeva, Sidford, Tian, and Zhao~\cite{jsstz}
introduced effective-resistance decompositions and obtained
$\breve O(n\log^2 n/\varepsilon^2)$-edge sparsifiers in nearly
linear time for polynomially bounded weights. Here $\breve O$ hides
factors polynomial in
$\log\log n$. They also derived faster Eulerian Laplacian solvers.
Zhao~\cite{zhao-dynamic} extended the study to fully dynamic
directed sparsification under a degree-balance-preserving
approximation. These results measure error through an undirected
Laplacian. Our result measures error in the two Schur-complement
metrics required by strong SV approximation. This distinction
matters when comparing support bounds and transferring guarantees
to products.

The recent Matrix Spencer theorem, proved concurrently by Akbas and Sra~\cite{as} and Song and Zhang \cite{song2026matrixspencerstandarddeviations},
and the simpler proof of Bansal and Kook~\cite{bk}, provide
further motivation for revisiting sparsification through
discrepancy. We use the variance-normalized Gaussian small-ball
estimate in~\cite[Theorem~1.4]{as}, together with the constrained
partial-coloring theorem of Reis and Rothvoss~\cite{rr}.
This theorem continues the line of constructive partial-coloring
methods started by Bansal~\cite{bansal} and by Lovett and
Meka~\cite{lm}, and extended to general convex bodies by
Rothvoss~\cite{rothvoss}. Reis and Rothvoss~\cite{rr-sparsifier} used
such methods to construct linear-size undirected spectral sparsifiers.
Our main structural contribution is the projected variance bound
that makes these tools apply in the strong-SV metrics.

The nearly linear construction also uses the Gaussian matrix bound
of Bandeira, Boedihardjo, and van Handel~\cite{bbvh}. Their estimate
depends on both matrix variance and the covariance operator of all
matrix entries. We bound the latter after extracting expander cores
with large minimum degree. This makes the expected Gaussian
discrepancy comparable to the square root of the projected variance.
A Gaussian rotation argument in the style of Maurey and
Pisier~\cite{pisier} transfers that estimate to
ordinary Euclidean projection onto the degree-constrained cube.

\subsection{Overview of the proof and algorithm}

We pass to the undirected bipartite lift of $A$ and apply the
effective-resistance decomposition of~\cite{jsstz}.
Each piece has small resistance diameter in the ambient graph.
We impose the degree equations separately on each piece and
project its normalized edge matrices onto the resulting subspace.

The key point is to keep track of the defect contributed by each
piece. Write $A_H$ for the adjacency supported on a piece $H$ and
$V_H^L$ for its left projected variance, as defined in
\cref{lem:er-variance}. That lemma bounds $V_H^L$ by
$\rho R_L(A)^{\dagger/2}R_L(A_H)R_L(A)^{\dagger/2}$, where
$\rho=O(n\log(2n)/m)$ at a stage with $m$ surviving arcs.
The Schur-complement inequality in \cref{lem:schur} then gives
\begin{align*}
 \sum_H V_H^L
 &\preceq \rho R_L(A)^{\dagger/2}
          \Bigl(\sum_H R_L(A_H)\Bigr)R_L(A)^{\dagger/2}\\
 &\preceq \rho R_L(A)^{\dagger/2}R_L(A)R_L(A)^{\dagger/2}
 \preceq \rho I.
\end{align*}
The same argument bounds the right projected variance.
The pieces may share vertices, but the sum incurs no factor for
that overlap. Overlap affects only the number of degree equations.
The Gaussian small-ball estimate and constrained partial coloring
now delete a constant fraction of the arcs with error
$O(\sqrt{n\log(2n)/m})$. Iterating to
$m=O(n\log(2n)/\varepsilon^2)$ gives the support bound.
Exact-degree rounding to a fixed grid controls the weight range
and the number of constraints throughout the iteration.

For the construction, we refine the resistance pieces into expanders
using~\cite{sw}. Centered edge matrices have a positive semidefinite
envelope of norm $\Lambda=\operatorname{polylog}n$, and sparse
products with approximate Laplacian inverses implement their action.
These ingredients first give a phase routine with time
$\softO((m+N)r^{-27/2})$ at radius $r\ge C\sqrt\rho$.
It already costs $\softO(m+N)$ when
$\rho>c\log^{-40}(2n)$.

For smaller $\rho$, we prune each expander to obtain large minimum
degree while keeping a constant fraction of the edges. If $\Pi$
projects onto the piecewise degree constraints and $g$ is a standard
Gaussian, the additional covariance bound gives
\begin{align*}
 \mathbb E\|\mathcal X(\Pi g)\|
 &\le C\left(\sqrt\rho+\rho\log^{23/2}(2n)\right)
 \le C'\sqrt\rho.
\end{align*}
Let $P$ be Euclidean projection onto the degree-constrained cube.
Oddness and nonexpansiveness of $P$, together with Gaussian rotation,
give
\begin{align*}
 \mathbb E\|\mathcal X(Pg)\|
 &\le\frac\pi2\mathbb E\|\mathcal X(\Pi g)\|
 \le C''\sqrt\rho.
\end{align*}
Thus ordinary cube projection gives the spectral guarantee with
constant probability. A sparse Jacobi iteration computes this
projection with logarithmic precision dependence. A spanning-tree
correction restores the degree equations exactly, and contraction
keeps the result in the cube. We check the discrepancy and saturation
count before accepting a trial.

In both branches, cycle rounding turns the nearly saturated
coordinates into zero-weight arcs. Its error is at most
$O(\beta\Lambda)$, where $\beta$ is the coordinate tolerance.
Resistance diameter controls the support bound, while expansion
and the split between the two routines control running time.
\Cref{sec:fast} gives the implementation, and
\cref{alg:fast-sv,alg:hybrid-phase} state the algorithm in pseudocode.

\section{SV approximation and the bipartite lift}
\label{sec:sv-lift}

For a nonnegative rectangular matrix $A\in\R^{r\times c}$, write
$D_L=\diag(A\1)$ and $D_R=\diag(A^\top\1)$ for its degree matrices.
Define its left defect $R_L(A)=D_L-AD_R^\dagger A^\top$ and its
right defect $R_R(A)=D_R-A^\top D_L^\dagger A$.
Here $\dagger$ denotes the Moore--Penrose pseudoinverse and
$R^{\dagger/2}$ its positive semidefinite square root.

\begin{definition}[Strong SV approximation]\label{def:sv}
A matrix $\widetilde A$ is an $\varepsilon$-SV approximation to $A$ if,
for all real $x\in\R^r$ and $y\in\R^c$,
\begin{equation}\label{eq:sv}
 |x^\top(\widetilde A-A)y|
 \le\varepsilon\sqrt{x^\top R_L(A)x}\sqrt{y^\top R_R(A)y}.
\end{equation}
We use the geometric-mean normalization of~\cite{lwz}. In the
normalization of~\cite{appsv}, the approximation parameter is
$2\varepsilon$.
\end{definition}

The inequality includes its kernel conditions: an error must vanish
on the left kernel of $R_L(A)$ and the right kernel of $R_R(A)$.
Together with those conditions, \eqref{eq:sv} is equivalent to
\begin{equation}\label{eq:sv-norm}
 \big\|R_L(A)^{\dagger/2}(\widetilde A-A)
                         R_R(A)^{\dagger/2}\big\|\le\varepsilon.
\end{equation}
The bilinear definition forces exact row and column sums since
$R_L(A)\1=R_R(A)\1=0$. A real operator-norm bound also gives the
corresponding complex-vector formulation.

We discard zero-degree rows and columns and set
$P=D_L^{-1/2}AD_R^{-1/2}$ on the remaining coordinates.
Weighted Cauchy--Schwarz gives
\begin{align*}
 \|Py\|_2^2
 &=\sum_u\frac1{(D_L)_{uu}}
       \left(\sum_v A_{uv}\frac{y_v}{\sqrt{(D_R)_{vv}}}\right)^2\\
 &\le\sum_{u,v}A_{uv}\frac{y_v^2}{(D_R)_{vv}}
 =\|y\|_2^2.
\end{align*}
Thus $P$ is a contraction. Write $F_L(P)=I-PP^\top$ and
$F_R(P)=I-P^\top P$ for its normalized defects. The identities
$R_L(A)=D_L^{1/2}F_L(P)D_L^{1/2}$ and
$R_R(A)=D_R^{1/2}F_R(P)D_R^{1/2}$ show that this normalization
preserves SV inequalities whenever the degrees stay fixed.

The undirected bipartite lift $G$ has a left vertex for each row, a
right vertex for each column, and an edge of conductance $A_{uv}$ for
each positive entry. Its Laplacian is
\[
 L_G=\begin{pmatrix}D_L&-A\\-A^\top&D_R\end{pmatrix}.
\]
The two defects are its Schur complements.
In particular,
\begin{equation}\label{eq:energy-min}
 x^\top R_L(A)x=\min_z\sum_{u,v}A_{uv}(x_u-z_v)^2.
\end{equation}
This identity proves positive semidefiniteness. It also identifies
$\ker R_L(A)$ with the vectors constant on the left vertices of each
connected component of $G$, and similarly on the right. Indeed, zero
energy in \eqref{eq:energy-min} forces equality along every lifted edge.

\begin{lemma}[Schur resistance and superadditivity]\label{lem:schur}
For two left vertices $u,a$ in one component of $G$,
\[
 (e_u-e_a)^\top R_L(A)^\dagger(e_u-e_a)=\operatorname{ER}_G(u,a),
\]
and analogously for two right vertices. If $A=\sum_iA_i$ with
$A_i\ge0$, then
\begin{equation}\label{eq:superadd}
 R_L(A)\succeq\sum_iR_L(A_i),
\end{equation}
and the same inequality holds with $R_R$ in place of $R_L$.
\end{lemma}
\begin{proof}
Let $b$ be a left-side demand whose entries sum to zero in each
component, and embed it in the lift as $\bar b=(b,0)$.
The variational formula for electrical energy and
\eqref{eq:energy-min} give
\begin{align*}
 \bar b^\top L_G^\dagger\bar b
 &=\sup_{x,z}\left\{2b^\top x-
             \begin{pmatrix}x\\z\end{pmatrix}^{\!\top}
             L_G\begin{pmatrix}x\\z\end{pmatrix}\right\}\\
 &=\sup_x\left\{2b^\top x-
                    \min_z\sum_{u,v}A_{uv}(x_u-z_v)^2\right\}\\
 &=\sup_x\{2b^\top x-x^\top R_L(A)x\}
 =b^\top R_L(A)^\dagger b.
\end{align*}
Taking $b=e_u-e_a$ proves the resistance identity. For
superadditivity, use \eqref{eq:energy-min} once more:
\begin{align*}
 x^\top R_L(A)x
 &=\min_z\sum_i\sum_{u,v}(A_i)_{uv}(x_u-z_v)^2\\
 &\ge\sum_i\min_z\sum_{u,v}(A_i)_{uv}(x_u-z_v)^2\\
 &=x^\top\left(\sum_iR_L(A_i)\right)x.
\end{align*}
Applying these calculations to $A^\top$ proves both right-side claims.
\end{proof}

The support bound in \cref{thm:main} is tight up to the logarithm.
Related lower bounds for undirected sparsifiers and cut sketches
appear in~\cite{bss,ckst}.

\begin{proposition}[A lower bound]\label{prop:lower}
Let $J_n$ be the $n\times n$ all-ones matrix and let $0<\varepsilon<1$.
Every $\varepsilon$-SV approximation $\widetilde A$ to $J_n$ has at
least
\[
 \frac{n^2}{1+\varepsilon^2n}
 \ge\frac12\min\left\{n^2,\frac n{\varepsilon^2}\right\}
\]
nonzero entries. This holds even if $\widetilde A$ may have negative
entries or arcs outside $\supp J_n$.
\end{proposition}
\begin{proof}
Both degree matrices of $J_n$ equal $nI$, so
$R_L(J_n)=R_R(J_n)=nI-J_n=n\Pi$, where $\Pi$ is the orthogonal
projection onto $\1^\perp$. The kernel conditions force
$\widetilde A\1=\widetilde A^\top\1=n\1$, so the error
$E=\widetilde A-J_n$ satisfies $E\1=E^\top\1=0$. Since
$R_L(J_n)^{\dagger/2}=R_R(J_n)^{\dagger/2}=n^{-1/2}\Pi$,
\eqref{eq:sv-norm} gives $\|E\|\le\varepsilon n$. The matrix $E$
annihilates $\1$, so its rank is at most $n-1$, and therefore
$\|E\|_F^2\le(n-1)\|E\|^2\le\varepsilon^2n^3$.

Let $s$ be the number of nonzero entries of $\widetilde A$. Its
entries sum to $n^2$, so the Cauchy--Schwarz inequality gives
$\sum_{u,v}\widetilde A_{uv}^2\ge n^4/s$. Expanding the Frobenius norm,
\begin{align*}
 \|E\|_F^2
 &=\sum_{u,v}\widetilde A_{uv}^2-2\sum_{u,v}\widetilde A_{uv}+n^2\\
 &=\sum_{u,v}\widetilde A_{uv}^2-n^2
 \ge\frac{n^4}s-n^2.
\end{align*}
Combining the two bounds gives $n^4/s\le n^2+\varepsilon^2n^3$, which
rearranges to $s\ge n^2/(1+\varepsilon^2n)$. If $\varepsilon^2n\le1$,
this is at least $n^2/2$, and otherwise it is at least
$n/(2\varepsilon^2)$.
\end{proof}

\section{Effective-resistance pieces and projected variance}

We use the following direct specialization of
\cite[Definition~9 and Proposition~10]{jsstz}, taking their weight-scale
parameter to be $2$.

\begin{theorem}[Effective-resistance decomposition]\label{thm:erdecomp}
Let $G$ be a weighted undirected graph with $N$ vertices, $m>0$ edges,
and positive-weight ratio at most $W$. There are edge-disjoint nonempty
pieces $H$ covering at least $m/2$ edges such that, with
$\rho=16N\log(N+1)/m$ and $J=\lceil\log_2W\rceil+3$,
every vertex belongs to at most $J$ pieces, each piece has weight ratio
at most $2$, and
\begin{equation}\label{eq:erdiam}
 \left(\max_{e\in E(H)}w_e\right)
 \max_{a,b\in V(H)}\operatorname{ER}_G(a,b)\le\rho.
\end{equation}
We measure these resistances in the ambient graph $G$.
\end{theorem}

Every piece lies in a single connected component, since effective
resistance between different components is infinite.

\begin{lemma}[Monotonicity under projection]\label{lem:proj}
For rectangular matrices $B_1,\ldots,B_k$ and orthogonal projections
$\Pi\preceq Q$ on $\R^k$, write
$B_i^\Pi=\sum_j\Pi_{ji}B_j$ and $B_i^Q=\sum_jQ_{ji}B_j$.
Then
\[
 \sum_iB_i^\Pi(B_i^\Pi)^\top
 \preceq\sum_iB_i^Q(B_i^Q)^\top,
\]
and the analogous inequality holds after transposing all matrices.
\end{lemma}
\begin{proof}
For a fixed vector $v$, form the matrix $Z$ whose columns are $B_j^\top v$.
Since $\Pi^2=\Pi$, $Q^2=Q$, and $Q-\Pi\succeq0$,
\begin{align*}
 v^\top\left(\sum_i B_i^\Pi(B_i^\Pi)^\top\right)v
 &=\sum_i\|(B_i^\Pi)^\top v\|_2^2
 =\|Z\Pi\|_F^2\\
 &=\tr(Z\Pi Z^\top)
 \le\tr(ZQZ^\top)
 =\|ZQ\|_F^2\\
 &=v^\top\left(\sum_i B_i^Q(B_i^Q)^\top\right)v.
\end{align*}
Repeating the calculation with $B_jv$ as the columns of $Z$ proves
the transposed inequality.
\end{proof}

\begin{lemma}[Piece-sensitive strong-SV variance]\label{lem:er-variance}
Let $G$ be the bipartite lift of $A$, and abbreviate
$R_L=R_L(A)$, $R_R=R_R(A)$. Let a nonempty piece $H$ satisfy
\eqref{eq:erdiam}, with left and right vertex sets $U_L,U_R$.
Let $\Pi_H$ be the Euclidean orthogonal projection in $\R^{E(H)}$ onto
\begin{equation}\label{eq:constraints}
 S_H=\left\{x:\ \sum_{e\ni v}w_ex_e=0
                         \text{ for every }v\in U_L\cup U_R\right\}.
\end{equation}
For each edge $e=(u,v)$ of $H$, define
$C_e=w_e R_L^{\dagger/2}e_ue_v^\top R_R^{\dagger/2}$ and its projection
$\widetilde C_e=\sum_f(\Pi_H)_{fe}C_f$.
Writing $A_H$ for the adjacency supported on this piece, we have
\begin{equation}\label{eq:left-piece-var}
 \sum_e\widetilde C_e\widetilde C_e^\top
 \preceq\rho R_L^{\dagger/2}R_L(A_H)R_L^{\dagger/2},
\end{equation}
and, on the right side,
\begin{equation}\label{eq:right-piece-var}
 \sum_e\widetilde C_e^\top\widetilde C_e
 \preceq\rho R_R^{\dagger/2}R_R(A_H)R_R^{\dagger/2}.
\end{equation}
\end{lemma}
\begin{proof}
We prove \eqref{eq:left-piece-var} first. Set $w_*=\max_{e\in H}w_e$,
$\mu_R=|U_R|^{-1}\1_{U_R}$, and
$b_v=R_R^{\dagger/2}(e_v-\mu_R)$, with all vectors in the full
right coordinate space. By \cref{lem:schur}, Jensen's inequality,
and \eqref{eq:erdiam},
\begin{align}
 \|b_v\|^2
 &=\left\|\frac1{|U_R|}\sum_{z\in U_R}
                      R_R^{\dagger/2}(e_v-e_z)\right\|^2\notag\\
 &\le\frac1{|U_R|}\sum_{z\in U_R}
                      \|R_R^{\dagger/2}(e_v-e_z)\|^2\notag\\
 &=\frac1{|U_R|}\sum_{z\in U_R}\operatorname{ER}_G(v,z)
 \le\frac\rho{w_*}.\label{eq:centered-right}
\end{align}
Define $C'_e=w_eR_L^{\dagger/2}e_ub_v^\top$. For every $x\in S_H$,
the weighted row equations give
\begin{align*}
 \sum_e x_e(C_e-C'_e)
 &=R_L^{\dagger/2}
      \left(\sum_u e_u\sum_{v:(u,v)\in H}w_{uv}x_{uv}\right)
                         \mu_R^\top R_R^{\dagger/2}\\
 &=0.
\end{align*}
We can therefore replace $C_e$ by $C'_e$ in every sum projected by
$\Pi_H$. Let $Q$ impose only the weighted column equations.
The inclusion $S_H\subseteq\operatorname{range}Q$ gives
$\Pi_H\preceq Q$, so \cref{lem:proj} bounds the required variance
by the variance of the $Q$-projected matrices $C'_e$.

The projection $Q$ acts separately on the edges into each right
vertex. Write $a_v$ for column $v$ of $A_H$, and define
$K(a)=\diag(a)-aa^\top/(\1^\top a)$ for nonzero $a\ge0$, with
$K(0)=0$. On the edges into $v$, the projection matrix is
$I-a_va_v^\top/\|a_v\|_2^2$, restricted to the positive coordinates
of $a_v$. Writing $a^{\circ2}$ for the coordinatewise square, we obtain
\begin{align*}
 \sum_{e:\,v(e)=v}\Big(\sum_fQ_{fe}C'_f\Big)
                           \Big(\sum_fQ_{fe}C'_f\Big)^\top
 &=\|b_v\|^2R_L^{\dagger/2}
       \left(\diag(a_v^{\circ2})
       -\frac{a_v^{\circ2}(a_v^{\circ2})^\top}{\|a_v\|_2^2}\right)
                           R_L^{\dagger/2}\\
 &=\|b_v\|^2R_L^{\dagger/2}K(a_v^{\circ2})R_L^{\dagger/2}.
\end{align*}
For any $z$, choose $t_*$ to minimize
$\sum_u a_{v,u}(z_u-t)^2$. Then
\begin{align*}
 z^\top K(a_v^{\circ2})z
 &=\min_t\sum_u a_{v,u}^2(z_u-t)^2\\
 &\le\sum_u a_{v,u}^2(z_u-t_*)^2
 \le w_*\sum_u a_{v,u}(z_u-t_*)^2\\
 &=w_*z^\top K(a_v)z.
\end{align*}
Combining this comparison with \eqref{eq:centered-right} gives
\begin{align*}
 \sum_e\widetilde C_e\widetilde C_e^\top
 &\preceq\sum_v\|b_v\|^2
            R_L^{\dagger/2}K(a_v^{\circ2})R_L^{\dagger/2}\\
 &\preceq\rho R_L^{\dagger/2}
                  \Big(\sum_vK(a_v)\Big)R_L^{\dagger/2}\\
 &=\rho R_L^{\dagger/2}R_L(A_H)R_L^{\dagger/2}.
\end{align*}
Applying the same argument to $A^\top$ proves
\eqref{eq:right-piece-var}.
\end{proof}

\begin{corollary}[Summing the pieces]\label{cor:total-variance}
For edge-disjoint pieces with common parameter $\rho$, let $\Pi$ be the
block-diagonal projection imposing \eqref{eq:constraints} separately
on each piece. The corresponding projected matrices satisfy
$\sum_e\widetilde C_e\widetilde C_e^\top\preceq\rho I_r$ and
$\sum_e\widetilde C_e^\top\widetilde C_e\preceq\rho I_c$.
Every unnormalized projected error annihilates the two defect kernels.
\end{corollary}
\begin{proof}
Write $A=A_{\rm rem}+\sum_H A_H$, where $A_{\rm rem}\ge0$
contains the uncovered edges. Summing \eqref{eq:left-piece-var}
and applying \cref{lem:schur} gives
\begin{align*}
 \sum_e\widetilde C_e\widetilde C_e^\top
 &\preceq\rho R_L^{\dagger/2}
                  \left(\sum_H R_L(A_H)\right)R_L^{\dagger/2}\\
 &\preceq\rho R_L^{\dagger/2}R_L(A)R_L^{\dagger/2}
 \preceq\rho I_r.
\end{align*}
The transposed calculation gives the right variance bound.
For the kernel conditions, write the unnormalized error as
$E=\sum_HE_H$. Each $E_H$ has zero margins and lies inside one
component of the lift. If $p\in\ker R_L(A)$ and $q\in\ker R_R(A)$,
let $p_H,q_H$ denote their constant values on the two sides of
that component. Then $p^\top E=\sum_H p_H\1^\top E_H=0$ and
$Eq=\sum_H q_HE_H\1=0$.
\end{proof}

\section{From projected variance to an exact constrained partial coloring}

We use two external theorems, stated here with their quantitative
parameters. The first is the matrix small-ball theorem of Akbas and Sra
\cite[Theorem~1.4]{as}: there are absolute constants
$C_{\rm sb},c_{\rm sb}>0$ such that, for any number $q$ of real symmetric
$d\times d$ matrices $B_1,\ldots,B_q$ satisfying
$\sum_iB_i^2\preceq I_d$, independent standard Gaussians satisfy
\begin{equation}\label{eq:projected-smallball}
 \Pr\left(\left\|\sum_i g_iB_i\right\|\le C_{\rm sb}\right)
 \ge e^{-c_{\rm sb}d}.
\end{equation}
In particular, the theorem does not assume $q\le d$ or $q=d$.
We need this Gaussian measure bound in its variance-normalized form.
A signing bound alone does not give the measure estimate used below.

The second is the constrained partial-coloring theorem of Reis and
Rothvoss \cite[Theorem~6]{rr}: for fixed
$\alpha,\beta,\tau>0$ there is an absolute dilation
$C_{\rm pc}=C_{\rm pc}(\alpha,\beta,\tau)$ such that, if a symmetric
convex set $K\subseteq\mathbb R^m$ satisfies
$\gamma_m(K)\ge e^{-\alpha m}$ and a linear subspace
$S\subseteq\mathbb R^m$ satisfies $\dim S\ge\beta m$, then
$C_{\rm pc}K\cap S\cap[-1,1]^m$ contains a vector with at least
$(\beta-\tau)m$ coordinates in $\{-1,1\}$.
Their theorem also gives a randomized polynomial-time construction in
its separation-oracle model.

\begin{lemma}[Projected-variance partial coloring]\label{lem:projected-pc}
Let $T:\mathbb R^m\longrightarrow\mathbb S^d$ be linear, let
$S\subseteq\mathbb R^m$ be a subspace, and let $P$ be the orthogonal
projection onto $S$. Suppose that $m\ge d$, that $\dim S\ge m/2$,
and that
\[
 \sum_{i=1}^m T(Pe_i)^2\preceq \rho I_d.
\]
There is an absolute constant $C$ and a vector
$x\in S\cap[-1,1]^m$ such that $\|T(x)\|\le C\sqrt\rho$ and
$\bigl|\{i:x_i=-1\}\bigr|\ge m/8$.
\end{lemma}

\begin{proof}
Assume first that $\rho>0$, and set $B_i=\rho^{-1/2}T(Pe_i)$, so
that $\sum_iB_i^2=\rho^{-1}\sum_iT(Pe_i)^2\preceq I_d$.
For a standard Gaussian $g\in\R^m$, linearity gives
$T(Pg)=\sum_i g_iT(Pe_i)=\sqrt\rho\sum_i g_iB_i$.
Define the symmetric convex cylinder
$K=\{z\in\R^m:\|T(Pz)\|\le C_{\rm sb}\sqrt\rho\}$.
Applying \eqref{eq:projected-smallball} in the ambient coefficient
space gives
\begin{align*}
 \gamma_m(K)
 &=\Pr\!\left[\left\|\sum_i g_iB_i\right\|\le C_{\rm sb}\right]\\
 &\ge e^{-c_{\rm sb}d}
 \ge e^{-c_{\rm sb}m}.
\end{align*}
The Reis--Rothvoss theorem, with
$\alpha=c_{\rm sb}$, $\beta=1/2$, and $\tau=1/4$, now gives
$x\in C_{\rm pc}K\cap S\cap[-1,1]^m$ with at least $m/4$
saturated coordinates. Choose between $x$ and $-x$ so that the
negative sign occurs more often among those coordinates. Symmetry
preserves feasibility, and $Px=x$ gives
$\|T(x)\|=\|T(Px)\|\le C_{\rm pc}C_{\rm sb}\sqrt\rho$.
The sign choice gives
$|\{i:x_i=-1\}|\ge\tfrac12|\{i:|x_i|=1\}|\ge m/8$.

If $\rho=0$, then for every vector $v$,
$\sum_i\|T(Pe_i)v\|_2^2\le0$. Hence each $T(Pe_i)=0$ and $T$
vanishes on $S$. Apply the partial-coloring theorem to $K=\R^m$
and make the same sign choice.

If the external theorem requires a bounded body, intersect $K$
with a Euclidean ball of radius $R$. For sufficiently large $R$,
this intersection has Gaussian measure at least
$\tfrac12e^{-c_{\rm sb}m}\ge e^{-(c_{\rm sb}+\log2)m}$.
Applying the theorem to this intersection changes only the absolute
dilation constant and still produces a point in the original cylinder.
\end{proof}

\Needspace{10\baselineskip}
\begin{lemma}[One sparsification phase]\label{lem:phase}
Let $A$ have $m$ positive entries and let its bipartite lift have $N\le2n$
nonisolated vertices and weight ratio at most $W$. Set
$J=\lceil\log_2W\rceil+3$. If
\begin{equation}\label{eq:phase-size}
 m\ge\max\{4NJ,2N\},
\end{equation}
there is a nonnegative matrix $A'$ supported on $A$, with the same
row and column sums, at most $15m/16$ positive entries, and strong SV
error relative to $A$ at most
\begin{equation}\label{eq:phase-error}
 \delta=C_{\rm ph}\sqrt{\frac{n\log(2n)}m},
\end{equation}
where $C_{\rm ph}$ is an absolute constant.
\end{lemma}
\begin{proof}
Apply \cref{thm:erdecomp}, and let $F$ be the $k\ge m/2$ covered
edges. We change only weights in $F$. The direct sum $S$ of the
spaces \eqref{eq:constraints} has codimension at most
$\sum_H|V(H)|\le NJ$. By \eqref{eq:phase-size}, we have
$k\ge m/2\ge N$ and $\dim S\ge k-NJ\ge k-m/4\ge k/2$.
Map each coefficient vector to the symmetric dilation of its
SV-normalized error,
\[
 T(x)=\begin{pmatrix}0&\sum_ex_eC_e\\
                         (\sum_ex_eC_e)^\top&0\end{pmatrix}.
\]
If $\Pi$ projects onto $S$, then \cref{cor:total-variance} gives
\[
 \sum_eT(\Pi e_e)^2
 =\begin{pmatrix}
       \sum_e\widetilde C_e\widetilde C_e^\top&0\\
       0&\sum_e\widetilde C_e^\top\widetilde C_e
    \end{pmatrix}
 \preceq\rho I_N,
\]
where $\rho=16N\log(N+1)/m\le64n\log(2n)/m$.
We can therefore apply \cref{lem:projected-pc} in $\R^k$ with
matrix dimension $N$.

Set $w'_e=w_e(1+x_e)$ on $F$ and keep all other weights fixed.
Because $x\in S\cap[-1,1]^F$, every new weight is nonnegative, and
every vertex $v$ satisfies
$\sum_{e\ni v}(w'_e-w_e)=\sum_{e\ni v}w_ex_e=0$.
Each coordinate with $x_e=-1$ deletes an arc, so
\[
 |\supp A'|\le m-|\{e\in F:x_e=-1\}|
              \le m-k/8\le15m/16.
\]
Finally,
\begin{align*}
 \|R_L^{\dagger/2}(A'-A)R_R^{\dagger/2}\|
 &=\left\|\sum_e x_eC_e\right\|
 =\|T(x)\|\\
 &\le C\sqrt\rho
 \le C_{\rm ph}\sqrt{\frac{n\log(2n)}m}.
\end{align*}
Together with the kernel conditions from \cref{cor:total-variance},
this proves the strong-SV bound.
\end{proof}

\section{Composing errors with fixed degrees}
\begin{lemma}[Stability of the SV defects]\label{lem:sv-defect-stability}
Let $P,\widehat P$ be real contractions of the same rectangular dimensions,
and write $F_L(P)=I-PP^\top$, $F_R(P)=I-P^\top P$.
Suppose, including the associated kernel conditions, that
\[
 |x^\top(\widehat P-P)y|
 \leq\delta\sqrt{x^\top F_L(P)x}\sqrt{y^\top F_R(P)y}.
\]
Then $F_L(\widehat P)\preceq(1+2\delta)F_L(P)$ and
$F_R(\widehat P)\preceq(1+2\delta)F_R(P)$.
These statements hold even when either defect is singular.
\end{lemma}
\begin{proof}
Set $E=\widehat P-P$ and $X=PP^\top$. Since $0\preceq X\preceq I$,
\begin{align*}
 PF_R(P)P^\top
 &=X-X^2\\
 &=I-X-(I-X)^2
 \preceq I-X=F_L(P).
\end{align*}
Apply the assumed bilinear bound with $y=P^\top x$:
\begin{align*}
 |x^\top EP^\top x|
 &\le\delta\sqrt{x^\top F_L(P)x}
          \sqrt{x^\top PF_R(P)P^\top x}\\
 &\le\delta x^\top F_L(P)x.
\end{align*}
Expanding the new defect now gives, for every $x$,
\begin{align*}
 x^\top F_L(\widehat P)x
 &=x^\top F_L(P)x-2x^\top EP^\top x-\|E^\top x\|_2^2\\
 &\le x^\top F_L(P)x+2|x^\top EP^\top x|\\
 &\le(1+2\delta)x^\top F_L(P)x.
\end{align*}
Apply the same argument to $P^\top,\widehat P^\top$ for the
right defect. Each calculation remains valid when the defects are
singular.
\end{proof}

\begin{lemma}[Composition with exact degrees]\label{lem:sv-composition}
Let $A_0,\ldots,A_T$ be nonnegative rectangular matrices with identical
row and column sums, and normalize all of them using these common
degrees to obtain contractions $P_0,\ldots,P_T$.
Suppose that for each $t=0,\ldots,T-1$ the increment can be written
$P_{t+1}-P_t=B_t+R_t$, where the first term is small in the
current metric,
\[
 |x^\top B_t y|\leq\delta_t
  \sqrt{x^\top F_L(P_t)x}\sqrt{y^\top F_R(P_t)y},
\]
and the second term is small in the initial metric,
\[
 |x^\top R_t y|\leq\alpha_t
  \sqrt{x^\top F_L(P_0)x}\sqrt{y^\top F_R(P_0)y}.
\]
Set $D=\sum_t\delta_t$ and $a=\sum_t\alpha_t$.
Then $A_T$ is a strong SV approximation to $A_0$ with error at most
\[
 \frac{e^{2D}-1}{2}+e^{2D}a
 \leq e^{2D}(D+a).
\]
In particular, if $0<\varepsilon<1$ and $D+a\leq\varepsilon/4$,
the final error is at most $\varepsilon$.
If $R_t=0$ for every $t$, the sharper bound
$\tfrac12\bigl(\prod_t(1+2\delta_t)-1\bigr)$ holds.
\end{lemma}
\begin{proof}
Discard the common zero-degree coordinates. Define
$\beta_0=0$ and
$\beta_{t+1}=(1+2\delta_t)\beta_t+\delta_t+\alpha_t$.
We prove by induction that $P_t-P_0$ has strong-SV error at most
$\beta_t$ relative to $P_0$. Let
$\Psi_0(x,y)=\sqrt{x^\top F_L(P_0)x}\sqrt{y^\top F_R(P_0)y}$.
The induction hypothesis and \cref{lem:sv-defect-stability} give
$F_L(P_t)\preceq(1+2\beta_t)F_L(P_0)$ and
$F_R(P_t)\preceq(1+2\beta_t)F_R(P_0)$.
We can therefore measure all three terms of the next increment
against $\Psi_0$:
\begin{align*}
 |x^\top(P_{t+1}-P_0)y|
 &\le |x^\top(P_t-P_0)y|+|x^\top B_ty|+|x^\top R_ty|\\
 &\le\bigl(\beta_t+\delta_t(1+2\beta_t)+\alpha_t\bigr)\Psi_0(x,y)\\
 &=\beta_{t+1}\Psi_0(x,y).
\end{align*}
This also proves the kernel conditions, since the right side
vanishes whenever either defect energy is zero.

Set $\theta_t=\beta_t+1/2$. The recurrence becomes
$\theta_{t+1}=(1+2\delta_t)\theta_t+\alpha_t$, with $\theta_0=1/2$.
Unrolling it and using $1+2\delta_t\le e^{2\delta_t}$ yields
\begin{align*}
 \beta_T
 &=\frac12\left(\prod_{t=0}^{T-1}(1+2\delta_t)-1\right)
   +\sum_{t=0}^{T-1}\alpha_t
             \prod_{s=t+1}^{T-1}(1+2\delta_s)\\
 &\le\frac{e^{2D}-1}{2}+e^{2D}a\\
 &=\int_0^D e^{2u}\,du+e^{2D}a
 \le e^{2D}(D+a).
\end{align*}
When every $\alpha_t=0$, the first line gives the sharper product
bound. If $D+a\le\varepsilon/4$, then
$e^{2D}(D+a)\le e^{\varepsilon/2}\varepsilon/4<\varepsilon$.
\end{proof}

\section{Exact-degree rounding and weight control}

\begin{lemma}[An entrywise bound in the original SV metric]
\label{lem:original-entrywise-sv}
Let $A_0\in\{0,1\}^{n\times n}$, and write
$S_0=R_L(A_0)$ and $T_0=R_R(A_0)$.
Suppose $E$ is supported on $A_0$ and satisfies
$E\mathbf1=E^\top\mathbf1=0$. Then, for every real $x,y$,
\[
 |x^\top Ey|\le (2n-1)\left(\sum_{u,v}|E_{uv}|\right)
            \sqrt{x^\top S_0x}\sqrt{y^\top T_0y}.
\]
In particular, $E$ obeys the kernel conditions of SV approximation and
its original-metric SV norm is at most
$(2n-1)\sum_{u,v}|E_{uv}|$.
\end{lemma}
\begin{proof}
Let $G_0$ be the bipartite lift of $A_0$. Any two vertices in the
same component have a path with at most $2n-1$ unit-resistance
edges, so their effective resistance is at most $2n-1$.
Choose reference vertices $a_K,b_K$ on the two sides of each
nontrivial component $K$. By \cref{lem:schur} and Cauchy--Schwarz,
\begin{align*}
 |x_u-x_{a_K}|
 &\le\|S_0^{\dagger/2}(e_u-e_{a_K})\|_2
                          \|S_0^{1/2}x\|_2\\
 &=\sqrt{\operatorname{ER}_{G_0}(u,a_K)}
                          \sqrt{x^\top S_0x}
 \le\sqrt{(2n-1)x^\top S_0x}.
\end{align*}
In the same way, $|y_v-y_{b_K}|\le\sqrt{(2n-1)y^\top T_0y}$.
The support condition keeps every nonzero $E_{uv}$ within one
component, and the zero margins cancel the reference values. Hence
\begin{align*}
 |x^\top Ey|
 &=\left|\sum_K\sum_{u,v\in K}
                  E_{uv}(x_u-x_{a_K})(y_v-y_{b_K})\right|\\
 &\le\sum_K\sum_{u,v\in K}
                  |E_{uv}|\,|x_u-x_{a_K}|\,|y_v-y_{b_K}|\\
 &\le(2n-1)\left(\sum_{u,v}|E_{uv}|\right)
                    \sqrt{x^\top S_0x}\sqrt{y^\top T_0y}.
\end{align*}
If either energy is zero, the same bound gives $x^\top Ey=0$,
which proves the kernel conditions.
\end{proof}

\Needspace{15\baselineskip}
\begin{lemma}[Degree-preserving rounding onto a fixed grid]
\label{lem:grid-rounding}
Let $A_0\in\{0,1\}^{n\times n}$ have $m_0\ge1$ nonzero entries, and let
$A\ge0$ be supported on $A_0$ with the same row and column sums.
Suppose that the SV error of $A$ relative to $A_0$, measured against
$S_0,T_0$ in Lemma~\ref{lem:original-entrywise-sv}, is at most $1/4$.
For an integer $K\ge4m_0^2$, set $q=1/K$. Maximum-spanning forests
and tree elimination produce a nonnegative matrix $\widehat A$
with the following properties:
\begin{enumerate}
\item $\operatorname{supp}\widehat A\subseteq\operatorname{supp}A$.
\item $\widehat A$ has exactly the same row and column sums as $A_0$.
\item every entry of $\widehat A$ is an integer multiple of $q$, and every
      positive entry belongs to $[q,n]$.
\item $\sum_{u,v}|\widehat A_{uv}-A_{uv}|\le2nm_0q$.
\item the SV error of $\widehat A-A$, measured against $S_0,T_0$, is at
      most $4n^2m_0q$.
\end{enumerate}
For rational input $A$, the construction uses polynomially many exact
rational operations with polynomial bit complexity.
\end{lemma}
\begin{proof}
Write $\mathcal A_0,\mathcal A$ for the adjacencies of the bipartite
lifts and $D$ for their common degree matrix. Discard isolated
vertices, and set $M=D^{-1/2}\mathcal A_0D^{-1/2}$ and $F=I-M^2$.
The identity $D^{1/2}FD^{1/2}=\diag(S_0,T_0)$ lets us compare
the lifted adjacency error with the defect. For $z=(x,y)$,
\begin{align*}
 |z^\top(\mathcal A-\mathcal A_0)z|
 &=2|x^\top(A-A_0)y|\\
 &\le\tfrac12\sqrt{x^\top S_0x}\sqrt{y^\top T_0y}\\
 &\le\tfrac14(x^\top S_0x+y^\top T_0y)
 =\tfrac14 z^\top D^{1/2}FD^{1/2}z.
\end{align*}
The spectrum of $M$ lies in $[-1,1]$, so
$F=(I-M)(I+M)\preceq2(I-M)$.
Writing $L_0,L$ for the lifted Laplacians, we obtain
$\pm(L-L_0)\preceq\tfrac14D^{1/2}FD^{1/2}\preceq\tfrac12L_0$, and
therefore $\tfrac12L_0\preceq L\preceq\tfrac32L_0$.
Thus $A$ and $A_0$ have the same nontrivial components. For every
cut $S$ that separates such a component,
$w_A(S,\bar S)\ge\tfrac12w_{A_0}(S,\bar S)\ge1/2$.

Choose a maximum-weight spanning tree in each current component,
and let $\mathcal T$ be their union. If $e\in\mathcal T$, let $S_e$
be one side of its fundamental cut. The tree exchange property
gives $w_f\le w_e$ for every current edge $f$ crossing that cut.
Consequently,
\[
 \tfrac12
 \le\sum_{f\in\delta(S_e)}w_f
 \le|\delta(S_e)|w_e\le m_0w_e,
\]
so every tree edge satisfies $w_e\ge1/(2m_0)$.

For every nontree edge $f$, round down to
$\widehat w_f=q\lfloor w_f/q\rfloor$ and write
$\delta_f=\widehat w_f-w_f\in(-q,0]$.
Orient all edges from left to right. Multiplying the right-side
degree equations by $-1$ turns degree preservation into the
incidence equation $B^\top\delta=0$.
The nontree changes create a zero-sum demand in each component,
so a unique tree flow $\delta_{\mathcal T}$ restores the degrees.
Across the fundamental cut of a tree edge $e$, this equation gives
$\delta_e=-\sum_{f\in\delta(S_e)\setminus\{e\}}\sigma_{ef}\delta_f$
for some signs $\sigma_{ef}\in\{-1,1\}$. Only nontree edges occur in
this sum, so
\[
 |\delta_e|
 \le\sum_{f\in\delta(S_e)\setminus\{e\}}|\delta_f|
 \le m_0q.
\]
Since $q\le1/(4m_0^2)$, every tree edge $e$ satisfies
\[
 \widehat w_e
 =w_e+\delta_e
 \ge\frac1{2m_0}-m_0q
 \ge\frac1{4m_0}>0.
\]
Nontree weights remain nonnegative as well, and we introduce no
new support edges.

To check the grid property, eliminate leaves of each tree.
The target degrees are integers, and all rounded nontree weights
lie in $q\mathbb Z$. At a leaf, the remaining tree weight equals
the residual degree minus the incident nontree weights, so it also
lies in $q\mathbb Z$. Removing that leaf preserves the same
property for the residual degrees. Induction therefore places every
tree weight on the grid. Every positive weight is at least $q$,
and degree preservation bounds it by an original degree, which
is at most $n$.

There are at most $m_0$ nontree edges and $2n-1$ tree edges. Thus
\begin{align*}
 \sum_e|\widehat w_e-w_e|
 &\le\sum_{e\notin\mathcal T}q+
                      \sum_{e\in\mathcal T}m_0q\\
 &\le m_0q+(2n-1)m_0q
 =2nm_0q.
\end{align*}
\Cref{lem:original-entrywise-sv} now gives the kernel conditions and
the SV bound
\begin{align*}
 \|S_0^{\dagger/2}(\widehat A-A)T_0^{\dagger/2}\|
 &\le(2n-1)\sum_e|\widehat w_e-w_e|\\
 &\le(2n-1)2nm_0q
 \le4n^2m_0q.
\end{align*}

We compute the forest, round the nontree weights, and solve the
tree equations in polynomial time. For rational input, leaf
elimination computes each final tree weight directly from integer
degrees and grid weights. These operations also give the claimed
polynomial bit complexity.
\end{proof}

\begin{corollary}[One grid for every sparsification phase]
\label{cor:one-grid}
For $0<\varepsilon\le1$, choose
\[
 K=\left\lceil\frac{128n^2m_0^2}{\varepsilon}\right\rceil.
\]
Consider a support-decreasing degree-preserving process, rounding after
every phase by Lemma~\ref{lem:grid-rounding}, and suppose the total
SV error immediately before each rounding is at most $1/4$. If each
phase removes at least one support edge, there are at most $m_0$ phases.
The sum of all rounding errors in the original SV metric is at most
$\varepsilon/32$. After every rounding, positive weights have ratio at
most $nK$. If $\varepsilon\ge n^{-1/2}$, then
$\log(nK)=O(\log(2n))$.
\end{corollary}
\begin{proof}
The chosen denominator satisfies $K\ge4m_0^2$. If there are
$T\le m_0$ phases, \cref{lem:grid-rounding} bounds each
original-metric rounding error by
$\alpha_t\le4n^2m_0/K\le\varepsilon/(32m_0)$. Summing over the phases
gives
\[
 \sum_{t=0}^{T-1}\alpha_t
 \le\frac{T\varepsilon}{32m_0}
 \le\frac{\varepsilon}{32}.
\]
Every positive rounded weight lies in $[1/K,n]$, so its ratio
to any other positive weight is at most $nK$.
When $\varepsilon\ge n^{-1/2}$, we also have $m_0\le n^2$ and
\begin{align*}
 \log(nK)
 &\le\log\bigl(129n^3m_0^2\varepsilon^{-1}\bigr)\\
 &\le\log129+\tfrac{15}{2}\log n
 =O(\log(2n)).
\end{align*}
\end{proof}

\begin{remark}[Weighted original inputs]
\label{rem:weighted-grid-rounding}
The same rounding argument applies when the positive original
weights lie in $[1,U]$. We can reach this normalization by dividing
all weights by their positive minimum. Each original edge then has
conductance at least one, and every nontrivial component cut has
weight at least one. The resistance bound $2n-1$, the tree-weight
lower bound $1/(2m_0)$, and the correction bound $m_0q$ all remain
valid.

Original degrees need not be integers, so leaf elimination need
place the tree weights on the grid. It still keeps every positive
weight in the required range. Each nontree weight satisfies
$\widehat w_f\in\{0\}\cup[q,nU]$, and each tree weight satisfies
$\widehat w_e\in[1/(4m_0),nU]\subseteq[q,nU]$. Hence the ratio of
the largest positive weight to the smallest is at most $nUK$.
Using the same $K=\lceil128n^2m_0^2/\varepsilon\rceil$ gives
the same total rounding error. For $\varepsilon\ge n^{-1/2}$,
we have $\log(nUK)=O(\log(2n)+\log U)$. With rational original
and current weights, the computation also has polynomial
bit complexity.
\end{remark}

\section{The complete iteration}

\begin{proof}[Proof of \cref{thm:main}]
Let $m_0=|\supp A_0|$. If $m_0=0$ or $n=1$, return $A_0$.
We also return $A_0$ when $\varepsilon<n^{-1/2}$, because
$m_0\le n^2<n/\varepsilon^2$. We can therefore assume
$n\ge2$ and $\varepsilon\ge n^{-1/2}$.

Fix the grid denominator $K=\lceil128n^2m_0^2/\varepsilon\rceil$, the
shrinkage factor $q_*=15/16$, and the stopping threshold
\[
 M=\left\lceil C_*\frac{n\log(2n)}{\varepsilon^2}\right\rceil.
\]
We choose the absolute constant $C_*$ below. If $m_0\le M$,
return $A_0$. Otherwise, while $m_t:=|\supp A_t|>M$, apply
\cref{lem:phase} to produce $A'_t$, then apply
\cref{lem:grid-rounding} on the fixed grid $1/K$ to produce
$A_{t+1}$. Both steps preserve the original degrees and support.
We check their hypotheses by induction.

Assume that all previous roundings met their hypotheses.
The positive weights initially equal one and lie in $[1/K,n]$
after each rounding, so the weight ratio satisfies $W_t\le nK$.
Since $m_0\le n^2$ and $\varepsilon^{-1}\le\sqrt n$,
\[
 J_t=\lceil\log_2W_t\rceil+3
    \le\lceil\log_2(nK)\rceil+3
    \le C_J\log(2n).
\]
For sufficiently large $C_*$, the inequality $m_t>M$ therefore
implies $m_t\ge\max\{4NJ_t,2N\}$. We can apply \cref{lem:phase}.
Rounding introduces no support edges, so every completed phase obeys
\begin{equation}\label{eq:support-shrink}
 m_{t+1}\le q_*m_t.
\end{equation}
In particular, there are at most $m_0$ phases.

To check the next rounding, consider a prefix of $s$ phases,
including its final partial-coloring step before rounding.
For $0\le t<s$, \eqref{eq:support-shrink} gives
$m_t\ge q_*^{-(s-1-t)}m_{s-1}>q_*^{-(s-1-t)}M$.
Writing $\Gamma=n\log(2n)$, we obtain
\begin{align}
 D_s:=\sum_{t=0}^{s-1}\delta_t
 &\le C_{\rm ph}\sum_{t=0}^{s-1}\sqrt{\frac{\Gamma}{m_t}}\notag\\
 &\le C_{\rm ph}\sqrt{\frac{\Gamma}{M}}
                            \sum_{j=0}^{s-1}q_*^{j/2}\notag\\
 &\le\frac{C_{\rm ph}}{1-\sqrt{q_*}}\sqrt{\frac{\Gamma}{M}}
 \le\frac{\varepsilon}{8}.
                                      \label{eq:prefix-delta}
\end{align}
The last inequality holds once
$C_*\ge64C_{\rm ph}^2/(1-\sqrt{q_*})^2$.
For each completed rounding, \cref{cor:one-grid} gives
$\alpha_t\le\varepsilon/(32m_0)$. Taking $\alpha_t=0$ for an
unrounded step, we have
$a_s:=\sum_{t<s}\alpha_t\le m_0\varepsilon/(32m_0)=\varepsilon/32$.
The composition lemma now bounds the original-metric error by
\begin{align}
 e^{2D_s}(D_s+a_s)
 &\le e^{\varepsilon/4}
              \left(\frac{\varepsilon}{8}+\frac{\varepsilon}{32}\right)
                  \notag\\
 &\le e^{1/4}\frac{5\varepsilon}{32}
 <\frac14.                            \label{eq:prefix-invariant}
\end{align}
This verifies the next rounding hypothesis using only the preceding
steps, which closes the induction. After the last rounding, the
same bound gives final error at most
$e^{1/4}5\varepsilon/32<\varepsilon$.

The process stops with at most $M$ arcs and never increases support.
If it performs a phase, each final weight has the form $j/K$ with
$0\le j\le nK$. Its numerator and denominator therefore require
$O(\log n+\log(1/\varepsilon))$ bits. If it performs no phase,
the output remains the zero-one input.
\end{proof}

\begin{proof}[Proof of \cref{thm:weighted}]
Return the input if it is empty. Otherwise, divide by its smallest
positive weight, so the positive entries lie in $[1,U]$. This
scaling preserves the SV parameter because
$R_L(cA)=cR_L(A)$ and $R_R(cA)=cR_R(A)$.
As before, return the input if $\varepsilon<n^{-1/2}$.
For the remaining case, keep the same grid denominator $K$
and choose
\[
 M=\left\lceil C_*\left(\frac{n\log(2n)}{\varepsilon^2}
                             +n\log(2U)\right)\right\rceil.
\]
Use the same iteration, with the weighted rounding procedure from
\cref{rem:weighted-grid-rounding}. At the start of a phase, the
weight ratio satisfies $W_t\le nUK$, so
\[
 J_t\le\lceil\log_2(nUK)\rceil+3
       \le C'_J\bigl(\log(2n)+\log(2U)\bigr).
\]
For sufficiently large $C_*$, this gives $m_t>M\ge\max\{4NJ_t,2N\}$.
The weight ratio affects this constraint count but does not enter
the resistance parameter. Thus each phase error obeys
$\delta_t\le C_{\rm ph}\sqrt{n\log(2n)/m_t}$, and each rounding error
obeys $\alpha_t\le\varepsilon/(32m_0)$. The geometric calculation in
\eqref{eq:prefix-delta} gives $\sum_{t<s}\delta_t\le\varepsilon/8$,
the rounding bounds give $\sum_{t<s}\alpha_t\le\varepsilon/32$, and
\eqref{eq:prefix-invariant} verifies each rounding.
Although tree weights need not lie on the grid, their lower bound
keeps the weight ratio controlled. The iteration therefore returns
an $\varepsilon$-SV approximation with at most $M$ arcs.
Multiplying its weights by the original minimum gives the desired
sparsifier of the original input.
\end{proof}

\begin{remark}[Role of the existence proof]
The iteration above proves the support and weight guarantees using
exact partial coloring. The next section replaces that step by fast
approximate optimization and degree-preserving cycle rounding, proving
\cref{thm:fast} for unweighted inputs. The weighted result in
\cref{thm:weighted} remains an existence result with a randomized
polynomial-time real-arithmetic construction. Output bit length and
the cost of a finite-precision implementation are separate questions.
\end{remark}

\section{A nearly-linear construction in real arithmetic}
\label{sec:fast}

This section proves \cref{thm:fast}. The algorithm repeatedly deletes
a constant fraction of the current arcs. Within each phase it first
finds pieces that have both small resistance diameter and inverse
polylogarithmic conductance. We first construct a spectral-projection
routine whose cost is polynomial in the inverse phase radius. We use
this routine only when that cost is polylogarithmic in $n$.
In denser phases, expander pruning and a Gaussian covariance estimate
allow projection onto the degree-constrained cube alone. A sparse
Jacobi iteration computes that projection with logarithmic precision
dependence. Both routines make many arc weights small, and cycle
rounding turns a constant fraction of them into zeros. Finally,
the fixed-grid procedure of \cref{lem:grid-rounding} controls the
weight range for the next phase. Edges outside the chosen pieces keep
their current weights during the phase.

The parameters have distinct purposes. The resistance parameter
$\rho$ determines a discrepancy radius of order $\sqrt\rho$.
The expander parameter bounds a PSD envelope by $\Lambda I$, with
$\Lambda=\operatorname{polylog}n$, and thereby controls the optimization
cost. We choose a tolerance $\beta$ so that the cycle rounding
error $\beta\Lambda$ is a small fraction of the discrepancy radius.
We implement all matrix operations implicitly, without forming
the degree-space projector or the dense edge matrices.

\begin{algorithm}[H]
\caption{\Call{FastSV}: strong-SV sparsification of an unweighted digraph}
\label{alg:fast-sv}
\small
\textbf{Input:} $A_0\in\{0,1\}^{n\times n}$, $0<\varepsilon<1$,
$0<\eta<1/2$.
We choose the constants $C_*,C_{\rm pc},c_{\rm del}>0$ in the proof.
\begin{enumerate}[label=\arabic*:,leftmargin=2em,itemsep=3pt]
\item $m_0\gets|\supp A_0|$.
      If $m_0=0$, $n=1$, or $\varepsilon<n^{-1/2}$, \textbf{return} $A_0$.
\item $K_g\gets\lceil128n^2m_0^2/\varepsilon\rceil$,
      $M\gets\lceil C_*n\log(2n)/\varepsilon^2\rceil$.
\item If $m_0\le M$, \textbf{return} $A_0$.
\item $\vartheta\gets1-c_{\rm del}$,
      $T_{\max}\gets1+\lceil\log(m_0)/(-\log\vartheta)\rceil$,
      $A\gets A_0$.
\item \textbf{while} $|\supp A|>M$ \textbf{do}
\item \hspace*{1em}$m\gets|\supp A|$,
      $N\gets$ number of nonisolated vertices of the bipartite lift of $A$.
\item \hspace*{1em}$\rho\gets16N\log(N+1)/m$,
      $r\gets C_{\rm pc}\sqrt\rho+\varepsilon/(128T_{\max})$.
\item \hspace*{1em}$A'\gets\Call{ImprovedPhase}(A,r,\eta/T_{\max})$
      using \cref{alg:hybrid-phase}.
      If the call reports failure, \textbf{return failure}.
\item \hspace*{1em}$A\gets\Call{GridRound}(A_0,A',K_g)$
      using \cref{lem:grid-rounding}.
\item \textbf{end while}
\item \textbf{return} $A$.
\end{enumerate}
\end{algorithm}

\subsection{Refining resistance pieces without increasing overlap}

For a weighted undirected graph $H$, its conductance is
\[
 \Phi(H)=\min_{\varnothing\ne S\subsetneq V(H)}
 \frac{w_H(S,V(H)\setminus S)}
      {\min\{\operatorname{vol}_H(S),
                    \operatorname{vol}_H(V(H)\setminus S)\}}.
\]
We remove isolated vertices throughout. The expander decomposition
algorithm of Saranurak and Wang~\cite{sw} partitions the vertices into
induced subgraphs of conductance at least $\phi$ in
$\softO(m/\phi)$ time. It cuts an $O(\phi\log^3(2n))$ fraction of
the total edge weight. When weights differ by at most two, the same
bound controls the fraction of edges cut, up to an absolute factor.

\begin{lemma}[Joint resistance and expansion guarantees]
\label{lem:joint-pieces}
Let the current bipartite lift have $N\le2n$ nonisolated vertices,
$m$ edges, and weight ratio $W$. Set $\rho=16N\log(N+1)/m$ and
$J=\lceil\log_2 W\rceil+3$ as in \cref{thm:erdecomp}, and set
$\phi=c_\phi\log^{-3}(2n)$, where $c_\phi>0$ is a sufficiently small
absolute constant.
A randomized algorithm finds edge-disjoint connected pieces
$\mathcal H$ in $\softO(m+N)$ time such that the pieces cover at
least $m/4$ edges, their vertex counts satisfy
$\sum_{H\in\mathcal H}|V(H)|\le NJ$, and every piece has conductance
$\Phi(H)\ge\phi$ and weight ratio at most $2$.
Every piece also satisfies \eqref{eq:erdiam} with the same $\rho$.
The logarithmic factors in the running time may depend on the requested
failure probability.
\end{lemma}
\begin{proof}
Apply \cref{thm:erdecomp}, and call its pieces $H_0$. In each $H_0$,
run the expander decomposition on the unweighted support with
conductance parameter $2\phi$. Choose $c_\phi$ small enough that
the decomposition cuts at most half the edges. Restoring the
original weights changes conductance by at most a factor of two:
\begin{align*}
 \Phi(H)
 &\ge\frac{\min_{e\in H_0}w_e}{\max_{e\in H_0}w_e}
                                      \Phi_{\rm unweighted}(H)\\
 &\ge\tfrac12(2\phi)=\phi.
\end{align*}
The refinement partitions the vertices of $H_0$ and keeps at
least half its edges. Summing over the original pieces gives
$\sum_H|E(H)|\ge\tfrac12\sum_{H_0}|E(H_0)|\ge m/4$ and
$\sum_H|V(H)|\le\sum_{H_0}|V(H_0)|\le NJ$.
For each refined piece $H\subseteq H_0$,
\begin{align*}
 \max_{e\in H}w_e\,\max_{u,v\in V(H)}\operatorname{ER}_G(u,v)
 &\le\max_{e\in H_0}w_e\,\max_{u,v\in V(H_0)}
                                      \operatorname{ER}_G(u,v)\\
 &\le\rho.
\end{align*}
The positive conductance makes every nonempty piece connected.
The original pieces are edge-disjoint, so their decomposition
times sum to $\softO(m/\phi)=\softO(m)$.
Adding the nearly linear resistance-decomposition cost~\cite{jsstz}
gives $\softO(m+N)$ in total.
\end{proof}

\subsection{A PSD envelope for the centered edge matrices}
\label{sec:envelope}

Fix the current adjacency $A$ and abbreviate $R_L=R_L(A)$ and
$R_R=R_R(A)$. For $H\in\mathcal H$, let $s_H=\1^\top A_H\1$ be its
total weight, and define its left and right centers
$\mu_L^H=A_H\1/s_H$ and $\mu_R^H=A_H^\top\1/s_H$.
The centers are degree-weighted probability vectors, embedded in the
full left and right coordinate spaces. For a left vertex $u$ and a
right vertex $v$ of $H$, define the centered vectors
$a_u^H=R_L^{\dagger/2}(e_u-\mu_L^H)$ and
$b_v^H=R_R^{\dagger/2}(e_v-\mu_R^H)$. For $e=(u,v)\in H$, define the
edge matrix
\begin{equation}\label{eq:fast-edge-matrices}
 X_e=w_e\begin{pmatrix}0&a_u^H(b_v^H)^\top\\
                         b_v^H(a_u^H)^\top&0\end{pmatrix}
\end{equation}
and its envelope
$D_e=w_e\diag\bigl(a_u^H(a_u^H)^\top,\,b_v^H(b_v^H)^\top\bigr)$.
Let $F=\bigcup_HE(H)$, $k=|F|$, and let $S\subseteq\R^F$ impose
\eqref{eq:constraints} on each refined piece. Write $\Pi$ for its
orthogonal projector and $\mathcal X(x)=\sum_{e\in F}x_eX_e$.

\begin{lemma}[Variance and envelope bounds]
\label{lem:fast-envelope}
For the pieces in \cref{lem:joint-pieces}, every edge satisfies
$-D_e\preceq X_e\preceq D_e$. The envelopes satisfy
\begin{equation}\label{eq:fast-envelope}
 \sum_{e\in F}D_e\preceq 2\phi^{-2}I,
\end{equation}
and the projected matrices satisfy
\begin{equation}\label{eq:fast-projected-variance}
 \sum_{e\in F}\mathcal X(\Pi e_e)^2\preceq\rho I.
\end{equation}
If $x\in S$, then $\|\mathcal X(x)\|$ is the strong-SV norm, relative
to $A$, of the adjacency perturbation $E_{uv}=w_{uv}x_{uv}$ on $F$.
This perturbation also satisfies the required kernel conditions.
\end{lemma}
\begin{proof}
For vectors $p,q$, the identity
\begin{align*}
 \begin{pmatrix}p\\q\end{pmatrix}^{\!\top}
 (D_e\pm X_e)
 \begin{pmatrix}p\\q\end{pmatrix}
 &=w_e\bigl((a_u^H)^\top p\pm(b_v^H)^\top q\bigr)^2
 \ge0
\end{align*}
proves the first assertion. To bound the sum, let
$d_L=A_H\1$, $d_R=A_H^\top\1$, and use only positive-degree
coordinates of $H$. The normalized bipartite adjacency
$P_H=D_L(H)^{-1/2}A_HD_R(H)^{-1/2}$ has leading singular value one.
Cheeger's inequality and bipartite spectral symmetry give
$\sigma_2(P_H)\le1-\phi^2/2$ on the complement of the leading
singular vectors. Consequently,
\begin{align}
 R_L(A_H)
 &=D_L(H)^{1/2}(I-P_HP_H^\top)D_L(H)^{1/2}\notag\\
 &\succeq\frac{\phi^2}{2}
       \left(D_L(H)-\frac{d_Ld_L^\top}{s_H}\right).
                                             \label{eq:expander-defect}
\end{align}
When a side has only one vertex, the corresponding centered subspace
is zero and the same inequality holds. Thus
\begin{align}
 \sum_{e=(u,v)\in H}w_ea_u^H(a_u^H)^\top
 &=R_L^{\dagger/2}
       \left(D_L(H)-\frac{d_Ld_L^\top}{s_H}\right)
                                      R_L^{\dagger/2}\notag\\
 &\preceq 2\phi^{-2}R_L^{\dagger/2}R_L(A_H)R_L^{\dagger/2}.
                                             \label{eq:piece-envelope}
\end{align}
Sum the left-block bounds and apply \eqref{eq:superadd}:
\begin{align*}
 \sum_H\sum_{e=(u,v)\in H}w_ea_u^H(a_u^H)^\top
 &\preceq2\phi^{-2}R_L^{\dagger/2}
                   \left(\sum_HR_L(A_H)\right)R_L^{\dagger/2}\\
 &\preceq2\phi^{-2}R_L^{\dagger/2}R_LR_L^{\dagger/2}
 \preceq2\phi^{-2}I.
\end{align*}
The transposed calculation bounds the right block by the same
matrix, proving \eqref{eq:fast-envelope}.

For $x\in S$, write $E_H=\sum_{e=(u,v)\in H}w_ex_ee_ue_v^\top$.
The constraints give $E_H\1=E_H^\top\1=0$, so
\begin{align}
 &\sum_{e=(u,v)\in H}w_ex_e(e_u-\mu_L^H)(e_v-\mu_R^H)^\top\notag\\
 &\quad=E_H-(E_H\1)(\mu_R^H)^\top
       -\mu_L^H(\1^\top E_H)
       +\mu_L^H(\1^\top E_H\1)(\mu_R^H)^\top\notag\\
 &\quad=E_H.
                                      \label{eq:double-center-cancel}
\end{align}
Applying this identity to each column of $\Pi$ identifies the
projected centered matrices with the projected matrices in
\cref{cor:total-variance}. Squaring their symmetric dilations proves
\eqref{eq:fast-projected-variance}. The same identity and that
corollary prove the norm and kernel statements.
\end{proof}

\subsection{Implicit normalization and paired PSD coefficients}

The square roots in the centered vectors $a_u^H,b_v^H$ are useful for
analysis. For computation we replace them by implicit rectangular
factors. Let $B_G$ be an edge-by-vertex oriented incidence matrix of
the current lift, let $W_G=\diag(w)$, and let $L_G=B_G^\top W_GB_G$.
Let $Q_L,Q_R$ project away the componentwise constant vectors on
their respective sides, and let $J_L,J_R$ be coordinate embeddings
into the lift. A Laplacian solver~\cite{st-solver} provides a fixed symmetric linear
operator $S_G$ with
\begin{equation}\label{eq:fixed-laplacian-inverse}
 (1-\xi)L_G^\dagger\preceq S_G\preceq(1+\xi)L_G^\dagger,
\end{equation}
where $\xi=1/10$.
We can construct this operator and apply it or its transpose in
$\softO(m+N)$ time~\cite[Proposition~7]{jrt}. Define the implicit
factors $Z_L=W_G^{1/2}B_GS_GJ_LQ_L$ and $Z_R=W_G^{1/2}B_GS_GJ_RQ_R$.

\begin{lemma}[Implicit factors and batched queries]
\label{lem:implicit-factors}
On the event \eqref{eq:fixed-laplacian-inverse},
\begin{equation}\label{eq:left-factor-gram}
 (1-\xi)^2R_L^\dagger
 \preceq Z_L^\top Z_L\preceq(1+\xi)^2R_L^\dagger,
\end{equation}
and the same bounds hold for $Z_R^\top Z_R$ with $R_R$ in place of
$R_L$.
Replace $a_u^H,b_v^H$ by
$\bar a_u^H=Z_L(e_u-\mu_L^H)$ and
$\bar b_v^H=Z_R(e_v-\mu_R^H)$ in \eqref{eq:fast-edge-matrices},
and denote the results by $\bar X_e,\bar D_e$.
For $\overline{\mathcal X}(x)=\sum_ex_e\bar X_e$ and
$\Lambda=8\phi^{-2}$, we have $\sum_e\bar D_e\preceq\Lambda I$ and
$\sum_e\overline{\mathcal X}(\Pi e_e)^2\preceq2\rho I$.
For $x\in S$, its actual strong-SV error is at most
$2\|\overline{\mathcal X}(x)\|$.

Define the vectors
\[
 v_e^\pm=\sqrt{\frac{w_e}{2\Lambda}}
            \begin{pmatrix}\bar a_u^H\\ \pm\bar b_v^H\end{pmatrix}
\]
and the rank-one PSD matrices $M_e^\pm=v_e^\pm(v_e^\pm)^\top$.
Their sum is at most $I$. A matrix-vector product with any linear
combination of the $M_e^\pm$, and the evaluation of all quadratic
forms $z^\top M_e^\pm z$ for one vector $z$, each cost
$\softO(m+N)$.
\end{lemma}
\begin{proof}
On $\operatorname{im}L_G$, the symmetric operator
$L_G^{1/2}S_GL_G^{1/2}$ has eigenvalues in $[1-\xi,1+\xi]$.
Squaring it and conjugating by $L_G^{\dagger/2}$ gives
$(1-\xi)^2L_G^\dagger\preceq S_GL_GS_G\preceq(1+\xi)^2L_G^\dagger$.
The zero-sum-demand identity in \cref{lem:schur} gives
$Q_LJ_L^\top L_G^\dagger J_LQ_L=R_L^\dagger$.
To obtain \eqref{eq:left-factor-gram}, use
$Z_L^\top Z_L=Q_LJ_L^\top S_GL_GS_GJ_LQ_L$ and conjugate the first
bound by $J_LQ_L$. Using $J_RQ_R$ instead gives the right-side
bounds. Hence
$Z_L=U_LR_L^{\dagger/2}$ and $Z_R=U_RR_R^{\dagger/2}$ for maps
whose singular values on the relevant images lie in
$[1-\xi,1+\xi]$. Set $U=\diag(U_L,U_R)$. Then
$\bar X_e=UX_eU^\top$ and $\bar D_e=UD_eU^\top$, so
\eqref{eq:fast-projected-variance} gives
\begin{align*}
 \sum_e\overline{\mathcal X}(\Pi e_e)^2
 &\preceq\|U\|^2U
       \left(\sum_e\mathcal X(\Pi e_e)^2\right)U^\top\\
 &\preceq(1+\xi)^4\rho I\preceq2\rho I.
\end{align*}
In the same way, \eqref{eq:fast-envelope} gives
$\sum_e\bar D_e\preceq2(1+\xi)^2\phi^{-2}I\preceq\Lambda I$.
On the two defect images, the lower singular-value bound gives
$\|\mathcal X(x)\|\le(1-\xi)^{-2}\|\overline{\mathcal X}(x)\|
\le2\|\overline{\mathcal X}(x)\|$.
\Cref{lem:fast-envelope} gives zero error on their kernels.

Let $V$ have columns $v_e^+,v_e^-$. To apply $V$, first aggregate
the edge coefficients at their endpoints and subtract the appropriate
piece centers, and then apply $Z_L,Z_R$. To apply $V^\top$, first
apply $Z_L^\top,Z_R^\top$, compute the inner product with each piece
center, and evaluate the two endpoint expressions for every edge.
The aggregation costs $O(k+\sum_H|V(H)|)$. In all nontrivial phases
this is $O(m)$, because $m\ge C NJ$. Thus both products cost
$\softO(m+N)$. Finally,
$\sum_{e,\pm}M_e^\pm=\Lambda^{-1}\sum_e\bar D_e\preceq I$, and for
coefficients $y$ and a vector $z$ we have
$\bigl(\sum_{e,\pm}y_e^\pm M_e^\pm\bigr)z=V\diag(y)V^\top z$ and
$\bigl(z^\top M_e^\pm z\bigr)_{e,\pm}=(V^\top z)^{\circ2}$.
These formulas establish the claimed batched costs without forming
any column of $V$ explicitly.
\end{proof}

Define the paired constraint space
$S_{\pm}=\{(x,-x):x\in S\}\subseteq\R^{2k}$ and the matrix map
$\mathcal M(y)=\sum_e(y_e^+M_e^++y_e^-M_e^-)$. In particular,
$\mathcal M(x,-x)=\Lambda^{-1}\overline{\mathcal X}(x)$.
We include $y^++y^-=0$ in the optimization domain, so every
feasible iterate preserves the pairing.

\begin{lemma}[Gaussian measure after pairing]
\label{lem:paired-gaussian}
If $m\ge C NJ$ for a sufficiently large absolute constant, then
$\dim S_{\pm}\ge k/2$. Let $\Pi_{\pm}$ be its orthogonal projector.
The coefficient matrices $\mathcal M(\Pi_{\pm}e_i)$ satisfy
\begin{equation}\label{eq:paired-variance}
 \sum_{i=1}^{2k}\mathcal M(\Pi_{\pm}e_i)^2
 \preceq\frac{\rho}{\Lambda^2}I.
\end{equation}
Their common image has dimension at most $N$. Consequently, for an
absolute constant $C$, the cylinder
\[
 \mathcal K=\{y:\|\mathcal M(\Pi_{\pm}y)\|
                         \le C\sqrt\rho/\Lambda\}
\]
has Gaussian measure at least $\exp(-O(N))$ in $\R^{2k}$.
\end{lemma}
\begin{proof}
The dimension bound follows from the number of degree constraints:
\begin{align*}
 \dim S_{\pm}
 &=\dim S\ge k-NJ
 \ge k-m/8\ge k/2,
\end{align*}
where we choose $C\ge8$ and use $k\ge m/4$.
Let $Qx=(x,-x)/\sqrt2$. Then $Q$ is an isometry,
$\Pi_{\pm}=Q\Pi Q^\top$, and
$\mathcal M(Qx)=\overline{\mathcal X}(x)/(\sqrt2\Lambda)$.
For the positive and negative coordinate copies $e_e^+,e_e^-$, this
gives $\mathcal M(\Pi_{\pm}e_e^\pm)
=\pm\frac1{2\Lambda}\overline{\mathcal X}(\Pi e_e)$. Therefore
\begin{align*}
 \sum_{i=1}^{2k}\mathcal M(\Pi_{\pm}e_i)^2
 &=\frac1{2\Lambda^2}
                   \sum_{e\in F}\overline{\mathcal X}(\Pi e_e)^2\\
 &\preceq\frac\rho{\Lambda^2}I.
\end{align*}
All these matrices act inside
$\operatorname{im}Z_L\oplus\operatorname{im}Z_R$, a subspace of
dimension $d_*\le N$. Apply \eqref{eq:projected-smallball} on
that subspace:
\begin{align*}
 \gamma_{2k}(\mathcal K)
 &=\Pr\!\left[
       \left\|\sum_{i=1}^{2k}g_i\mathcal M(\Pi_{\pm}e_i)\right\|
                           \le C\sqrt\rho/\Lambda\right]\\
 &\ge e^{-c_{\rm sb}d_*}
 \ge e^{-c_{\rm sb}N}.
\end{align*}
Thus the measure estimate depends on the common image dimension,
even though we implement the matrices in dimension $2m$.
\end{proof}

\subsection{Fast optimization with exact degree constraints}

We now implement the optimization step. Every iterate satisfies
the linear constraints exactly. We approximate the objective value
and the distance to the Gaussian vector.

\begin{lemma}[Linear optimization on the paired space]
\label{lem:fast-linear}
For every $0<\Delta<1$ and objective $c\in\R^{2k}$, we can find
$y\in S_{\pm}\cap[-1,1]^{2k}$ with
\[
 c^\top y\le(1-\Delta)
       \min_{z\in S_{\pm}\cap[-1,1]^{2k}}c^\top z
\]
in $\softO((m+N)\Delta^{-2})$ real-arithmetic operations.
\end{lemma}
\begin{proof}
Writing $y=(x,-x)$ changes the objective to $(c^+-c^-)^\top x$.
Make a separate copy of every vertex in each piece, and orient all
edges from left to right. The constraint on $x$ is then
$B^\top Wx=0$ for the disjoint union of the pieces. Since each
piece is bipartite, this equation preserves every unsigned degree.
We use the linear optimization oracle of~\cite[Lemma~15]{jrt},
which requires an exact routing map with bounded congestion.

In a connected piece, write $L=B^\top WB=D-\mathcal A$ and set
$P=\tfrac12(I+D^{-1}\mathcal A)$. For
$t=O(\phi^{-2}\log(2m))$, define
$H_t=\tfrac12\sum_{j=0}^{t-1}P^jD^{-1}$.
Let $R_T$ route zero-sum demands on a spanning tree. We use
\begin{equation}\label{eq:exact-routing-map}
 R=WBH_t+R_T(I-LH_t).
\end{equation}
Because $L=2D(I-P)$, every zero-sum demand $b$ satisfies
\begin{align*}
 LH_tb
 &=D(I-P)\sum_{j=0}^{t-1}P^jD^{-1}b\\
 &=D(I-P^t)D^{-1}b
 =b-DP^tD^{-1}b.
\end{align*}
The tree routing $R_T$ routes the zero-sum demand $(I-LH_t)b$
exactly, so $B^\top Rb=LH_tb+(I-LH_t)b=b$. Thus the tree correction
enforces exact conservation.

To bound congestion, take $b=B^\top Wz$ with $\|z\|_\infty\le1$
and let $f=D^{-1}b$. Then $\|f\|_\infty\le1$ and $f$ has
stationary mean zero. Write $\pi$ for the lazy walk's stationary
distribution and $w_{\min}$ for the piece's smallest weight.
The conductance bound and weight ratio give a spectral gap of at
least $\phi^2/4$, $\pi_{\min}\ge1/(4m)$, and
$\operatorname{vol}(H)\le4mw_{\min}$. Therefore
\begin{align*}
 \|P^tf\|_\infty
 &\le\pi_{\min}^{-1/2}\|P^tf\|_{2,\pi}\\
 &\le2\sqrt m\,(1-\phi^2/4)^t\|f\|_{2,\pi}
 \le2\sqrt m\,e^{-t\phi^2/4},
\end{align*}
and hence
\begin{align*}
 \|DP^tf\|_1
 &\le\operatorname{vol}(H)\|P^tf\|_\infty\\
 &\le8m^{3/2}w_{\min}e^{-t\phi^2/4}
 \le w_{\min}/2.
\end{align*}
We ensure the last inequality by choosing the constant in $t$
sufficiently large. A tree edge carries a signed sum of a subset
of the demands, so $\|W^{-1}R_TDP^tf\|_\infty\le\|DP^tf\|_1/w_{\min}\le1$.
Each walk term has
$\|BP^jf\|_\infty\le2\|P^jf\|_\infty\le2$, so
$\|BH_tb\|_\infty\le t$. Since $Rb=WBH_tb+R_TDP^tf$, we conclude
\[
 \|W^{-1}Rb\|_\infty
 \le\|BH_tb\|_\infty+\|W^{-1}R_TDP^tf\|_\infty
 \le t+1.
\]
Sparse walk steps and tree traversals apply $R$ or $R^\top$ in
$O(t|E(H)|)$ operations. The oracle of~\cite[Lemma~15]{jrt}
uses $O((t+1)^2\log(2k)/\Delta^2)$ such applications. Since
the pieces are edge-disjoint and $\phi^{-1}=\operatorname{polylog}n$,
the total cost is $\softO((m+N)\Delta^{-2})$.
\end{proof}

We record the approximate-projection interface with its parameter
dependence. This also explains why the dense appearance of the
normalized matrices does not cause a quadratic implementation cost.

\begin{lemma}[Approximate Gaussian projection]
\label{lem:fast-projection}
Let $q=2k$, $0<\sigma,\beta<1$, and $g\in\R^q$ satisfy
$\|g\|_2\le2\sqrt q$. Fix an absolute constant $c_0>0$.
With any prescribed failure probability $\zeta>0$, we can compute
$y\in S_{\pm}\cap[-1,1]^q$ with $\|\mathcal M(y)\|\le\sigma$ and
\begin{equation}\label{eq:approx-proj-objective}
 \|y-g\|_2^2\le
 \min_{\substack{z\in S_{\pm}\cap[-1,1]^q\\
                         \|\mathcal M(z)\|\le\sigma}}
             \|z-g\|_2^2+c_0q\beta^2/4.
\end{equation}
The running time is
\begin{equation}\label{eq:approx-proj-time}
 \softO\!\left((m+N)
       \bigl(\sigma^{-9/2}\beta^{-9}
                       +\sigma^{-2}\beta^{-8}\bigr)\right),
\end{equation}
where logarithmic factors include $\log(1/\zeta)$.
\end{lemma}
\begin{proof}
Apply the regularization and binary-search reduction
of~\cite[Proposition~2 and Section~7.1]{jrt} on
$\mathcal D=S_{\pm}\cap[-1,1]^q$, with
$f(y)=\|\mathcal M(y)\|$ and
$f_\lambda(y)=f(y)+\lambda\|y-g\|_2^2$.
The domain is symmetric and convex, contains zero, and
$f(0)=0$. For $y$ in the cube, positivity of the coefficients gives
$\pm\mathcal M(y)\preceq\sum_{e,\pm}M_e^\pm\preceq I$,
so $f(y)\le1$.
The reduction solves a logarithmic number of regularized problems
with $\lambda\ge\sigma/(8q)$ and additive objective accuracy
$\Theta(\lambda q\beta^2)$. It estimates operator norms to
multiplicative accuracy $1+O(\beta^2)$.

Set $\mathcal B(y)=\diag(\mathcal M(y),-\mathcal M(y))$,
and let $d$ be its dimension. We smooth $f$ with the function
$\mu\log\tr\exp(\mathcal B(y)/\mu)$ and choose
$\mu=\Theta(\lambda q\beta^2/\log(2d))$.
The Frank--Wolfe analysis in~\cite[Lemma~12]{jrt} uses
$N_{\rm FW}=\softO(\sigma^{-2}\beta^{-4})$ iterations, linear-oracle
accuracy $\Delta_{\rm lin}=\Theta(\beta^2)$, and gradient error
$\Delta_{\rm grad}=\Theta(\sigma\beta^2)$ in $\ell_1$. Throughout the
regularization interval, the smoothing parameter satisfies
$\mu^{-1}=\softO(\sigma^{-1}\beta^{-2})$.

The Gibbs-gradient estimator in~\cite[Lemma~11]{jrt} uses
polynomial matrix-vector products and batched quadratic forms.
We evaluate all forms $z^\top M_e^\pm z$ together as
$(V^\top z)^{\circ2}$, so \cref{lem:implicit-factors} bounds
the cost per batch by $\softO(m+N)$. Each gradient estimate
therefore costs
$T_{\rm grad}=\softO\bigl((m+N)\mu^{-1/2}\Delta_{\rm grad}^{-2}\bigr)
=\softO\bigl((m+N)\sigma^{-5/2}\beta^{-5}\bigr)$, and by
\cref{lem:fast-linear} each linear-oracle call costs
$T_{\rm lin}=\softO\bigl((m+N)\Delta_{\rm lin}^{-2}\bigr)
=\softO\bigl((m+N)\beta^{-4}\bigr)$. The total cost is
\[
 N_{\rm FW}(T_{\rm grad}+T_{\rm lin})
 =\softO\!\left((m+N)
       \bigl(\sigma^{-9/2}\beta^{-9}
                   +\sigma^{-2}\beta^{-8}\bigr)\right).
\]
Norm estimation and the logarithmic number of regularization
queries fit within the same bound.

To preserve the homogeneous constraints, initialize Frank--Wolfe
at zero and use its linear oracle on $\mathcal D$.
Every update is a convex combination of feasible points.
The binary-search reduction also uses only convex combinations
and scalar contractions. At its upper endpoint, minimize
$\|y-g\|_2^2$ approximately over $\mathcal D$ with the same
linear oracle. This replaces coordinatewise clipping, which would
not preserve the constraints. Additive accuracy
$\Theta(q\beta^2)$ takes $O(\beta^{-2})$ Frank--Wolfe iterations.
Their cost is $\softO((m+N)\beta^{-6})$, which the displayed bound
absorbs. Decreasing the accuracy constants by an absolute factor
leaves enough room for this error in
\eqref{eq:approx-proj-objective}. The final scalar contraction
enforces $\|\mathcal M(y)\|\le\sigma$.

Let $Q_{\max}$ bound the number of randomized queries, and give
each query conditional failure probability at most $\zeta/Q_{\max}$.
If $E_j$ is the event that query $j$ is the first inaccurate one,
then
\begin{align*}
 \Pr[\text{some query is inaccurate}]
 &=\sum_{j=1}^{Q_{\max}}\Pr[E_j]\\
 &\le\sum_{j=1}^{Q_{\max}}\frac{\zeta}{Q_{\max}}
 =\zeta.
\end{align*}
The extra dependence on $Q_{\max}/\zeta$ is logarithmic.
The linear equalities hold for every iterate, independently of
the objective estimates.
\end{proof}

\begin{lemma}[Fast almost-partial coloring]
\label{lem:fast-coloring}
There are absolute constants $C_{\rm pc},\gamma>0$ such that,
if $m\ge C NJ$ and
$C_{\rm pc}\sqrt\rho\le r\le1/8$, then with $\beta=r/(8\Lambda)$ we
can find $x\in S\cap[-1,1]^F$ satisfying
$\|\overline{\mathcal X}(x)\|\le r$ and
$|\{e:x_e\le-1+\beta\}|\ge\gamma k$.
The running time is $\softO((m+N)(\Lambda/r)^{27/2})$,
with logarithmic dependence on the desired failure probability.
\end{lemma}
\begin{proof}
Set $q=2k$ and $\sigma=r/\Lambda$.
By \cref{lem:paired-gaussian}, the projected discrepancy cylinder
has Gaussian measure at least $\exp(-O(N))\ge\exp(-O(q))$,
and $\dim S_{\pm}\ge q/4$. The constrained Gaussian projection
theorem~\cite[Theorem~6]{rr} therefore gives constants $c_0>0$
and $C_{\rm pc}$ such that the nearest point $y_*$ to a standard
Gaussian $g$ in
$\mathcal D_\sigma=\{y\in S_{\pm}\cap[-1,1]^q:
\|\mathcal M(y)\|\le\sigma\}$
has at least $c_0q$ coordinates of magnitude one with probability
$1-\exp(-\Omega(q))$. We choose $c_0\le1/8$ and take
$C_{\rm pc}$ large enough for the theorem's dilation.
We apply the measure theorem to the projected body.
On $S_{\pm}$ its matrix map equals $\mathcal M$, so its feasible
set is exactly $\mathcal D_\sigma$.

Apply \cref{lem:fast-projection} with $\beta=\sigma/8$.
The projection inequality
$\langle y-y_*,y_*-g\rangle\ge0$ gives
\[
 \|y-y_*\|_2^2
 \le\|y-g\|_2^2-\|y_*-g\|_2^2
 \le c_0q\beta^2/4.
\]
Hence at most $\beta^{-2}\|y-y_*\|_2^2\le c_0q/4$ coordinates satisfy
$|y_i-(y_*)_i|>\beta$, and therefore
$|\{i:|y_i|\ge1-\beta\}|\ge|\{i:|(y_*)_i|=1\}|-c_0q/4\ge c_0q/2$.
Since $y=(x,-x)$, each coordinate of $x$ contributes two
coordinates of $y$. Choose the sign of $x$ to maximize the
number near $-1$, and fix $\gamma=c_0/8$. Then
\begin{align*}
 |\{e:x_e\le-1+\beta\}|
 &\ge\tfrac14|\{i:|y_i|\ge1-\beta\}|\\
 &\ge c_0q/8=c_0k/4\ge\gamma k,
\end{align*}
and $\|\overline{\mathcal X}(x)\|=\Lambda\|\mathcal M(y)\|
\le\Lambda\sigma=r$.

Call a Gaussian good if $\|g\|_2\le2\sqrt q$ and its exact
projection has at least $c_0q$ saturated coordinates.
Gaussian concentration and the projection theorem give a fixed
lower bound $p_0>0$ on this event. Draw independently for at most
$L_{\rm try}=\lceil p_0^{-1}\log(2/\zeta)\rceil$ attempts, and
run each approximate projection with failure probability
$\zeta/(2L_{\rm try})$. If all projection queries succeed,
a good Gaussian produces an accepted vector. Therefore
\begin{align*}
 \Pr[\text{no valid vector is returned}]
 &\le\Pr[\text{all }L_{\rm try}\text{ Gaussians are bad}]
       +\Pr[\text{some projection query fails}]\\
 &\le(1-p_0)^{L_{\rm try}}
                +L_{\rm try}\frac{\zeta}{2L_{\rm try}}\\
 &\le e^{-p_0L_{\rm try}}+\zeta/2
 \le\zeta.
\end{align*}
Finally, substitute $\beta=\sigma/8$ in
\eqref{eq:approx-proj-time}. Including the trials costs
\begin{align*}
 \softO\bigl((m+N)(\sigma^{-27/2}+\sigma^{-10})\bigr)
 &=\softO\bigl((m+N)\sigma^{-27/2}\bigr)\\
 &=\softO\bigl((m+N)(\Lambda/r)^{27/2}\bigr).
\end{align*}
\end{proof}

\subsection{Deleting arcs by cycle rounding}

An almost-partial coloring is sufficient because the coordinates
near $-1$ already represent very small arc weights. We round those
weights using cycles within each piece, preserving its degrees.

\begin{lemma}[Rounding the small weights]
\label{lem:fast-cycle}
Let $x\in S\cap[-1,1]^F$ and $\beta>0$, and first replace $w_e$ by
$z_e=w_e(1+x_e)$ on $F$. In $O(m\log(2n))$ time, we can round
these weights while preserving each piece's degrees so that
at least
\[
 \frac{|\{e:x_e\le-1+\beta\}|-NJ}{2}
\]
of the original edges have weight zero. The additional strong-SV
error, in the metric of the graph at the beginning of the phase,
is at most $2\beta\Lambda$.
The same conclusion holds for any family of subpieces with total
vertex overlap at most $NJ$ and an implicit PSD envelope bounded
by $\Lambda I$.
\end{lemma}
\begin{proof}
In a piece $H$, let $L_H=\{e\in E(H):x_e\le-1+\beta\}$.
The weights on this set satisfy $0\le z_e\le\beta w_e$.
Keep all other weights fixed. Within $L_H$, alternately add and
subtract weight along an even cycle until an edge reaches an
endpoint of its interval $[0,2z_e]$, then remove that edge from
the active set. Every such update preserves the degrees.
When the active edges form a forest, at most $|V(H)|-1$ weights
remain strictly between their endpoints.

Let $B_H$ be an oriented incidence matrix for this subgraph,
and let $\hat z$ be the resulting weights, so that
$B_H^\top\hat z=B_H^\top z$. Reflection also preserves the degrees,
since $B_H^\top(2z-\hat z)=2B_H^\top z-B_H^\top\hat z=B_H^\top z$.
Each endpoint weight becomes zero in at least one of $\hat z$
and $2z-\hat z$. Choose the vector with more zeros. If $Z_H$
denotes its number of zero weights, then
$Z_H\ge\tfrac12\bigl(|L_H|-|V(H)|\bigr)$. Summing over the pieces gives
\begin{align*}
 \sum_H Z_H
 &\ge\tfrac12\left(\sum_H|L_H|-\sum_H|V(H)|\right)\\
 &\ge\tfrac12\bigl(|\{e:x_e\le-1+\beta\}|-NJ\bigr).
\end{align*}
Weights that already equal zero remain zero throughout.
The link/cut-tree implementation in~\cite[Lemma~16]{jrt} takes
$O(|L_H|\log(2n))$ time per piece, or $O(m\log(2n))$ in total.

Write the additional adjacency change as $\Delta A$, and set
$h_e=(\hat z_e-z_e)/w_e$ on the rounded edges and $h_e=0$
elsewhere. Degree preservation gives $h\in S$. Moreover,
$|h_e|\le z_e/w_e\le\beta$, so
$-\bar D_e\preceq\bar X_e\preceq\bar D_e$ gives
$-\beta\sum_e\bar D_e\preceq\overline{\mathcal X}(h)
\preceq\beta\sum_e\bar D_e$. \Cref{lem:implicit-factors} now gives
\[
 \|R_L^{\dagger/2}\Delta A R_R^{\dagger/2}\|
 \le2\|\overline{\mathcal X}(h)\|
 \le2\beta\Lambda.
\]
The zero margins within each piece also give the kernel
conditions by \cref{lem:fast-envelope}.
\end{proof}

\begin{algorithm}[H]
\caption{\Call{SpectralPhase}: a phase with polynomial inverse-radius cost}
\label{alg:fast-phase}
\small
\textbf{Input:} current adjacency $A$, radius $r$, failure probability $\zeta$.
\begin{enumerate}[label=\arabic*:,leftmargin=2em,itemsep=3pt]
\item Find the joint resistance/expander pieces $\mathcal H$
      of \cref{lem:joint-pieces}. Let $F=\bigcup_HE(H)$.
\item Build the implicit factors and paired PSD map of
      \cref{lem:implicit-factors}. Set
      $\Lambda=8\phi^{-2}$, $\sigma=r/\Lambda$, $\beta=\sigma/8$.
\item \textbf{repeat} at most $C\log(1/\zeta)$ times:
\item \hspace*{1em}Draw $g\sim\mathcal N(0,I_{2|F|})$.
      If $\|g\|_2>2\sqrt{2|F|}$, skip this attempt.
\item \hspace*{1em}Approximately project $g$ onto
      $S_{\pm}\cap[-1,1]^{2|F|}\cap\{y:\|\mathcal M(y)\|\le\sigma\}$
      using \cref{lem:fast-projection}, with tolerance $\beta$.
\item \hspace*{1em}Write the result as $(x,-x)$ and choose its sign
      to maximize $|\{e:x_e\le-1+\beta\}|$.
      If this count is at least $\gamma|F|$, exit the repeat loop.
\item \textbf{end repeat}. If no attempt meets the count,
      \textbf{return failure}.
\item Set $z_e\gets w_e(1+x_e)$ for $e\in F$ and keep
      $z_e\gets w_e$ for $e\notin F$.
\item For each $H$, apply the cycle rounding of
      \cref{lem:fast-cycle} to $\{e\in E(H):x_e\le-1+\beta\}$,
      with bounds $0\le\hat z_e\le2z_e$, keeping other edges fixed.
\item Delete zero-weight arcs and \textbf{return} the resulting $A'$.
\end{enumerate}
Run the decomposition, the implicit-factor construction, and the
coloring routine with failure probability $\zeta/3$ each.
\end{algorithm}

\begin{lemma}[Spectral-projection phase]
\label{lem:fast-phase}
There are absolute constants $C_0,C_{\rm pc},c_{\rm del}>0$ such
that, whenever $m\ge C_0NJ$ and
$C_{\rm pc}\sqrt\rho\le r\le1/8$, \cref{alg:fast-phase} returns,
with probability at least $1-\zeta$, a nonnegative reweighting $A'$
with the same row and column sums as $A$ and
$|\supp A'|\le(1-c_{\rm del})m$, whose error satisfies
\[
 \|R_L(A)^{\dagger/2}(A'-A)R_R(A)^{\dagger/2}\|\le4r
\]
and the corresponding strong-SV kernel conditions.
Its time is $\softO((m+N)r^{-27/2})$.
\end{lemma}
\begin{proof}
Since $k\ge m/4$, choosing $C_0\ge8/\gamma$ gives
$NJ\le\gamma m/8\le\gamma k/2$.
Let $\delta_{\rm phase}$ denote the total strong-SV error.
The coloring and cycle-rounding bounds give
\[
 m-|\supp A'|
 \ge\frac{\gamma k-NJ}{2}
 \ge\frac{\gamma k}{4}
 \ge\frac{\gamma m}{16}
\]
and $\delta_{\rm phase}\le2r+2\beta\Lambda=2r+r/4\le4r$.
Set $c_{\rm del}=\gamma/16$. Both updates preserve each piece's
degrees and leave the other weights fixed. Their kernel
conditions therefore hold for the total change.

The decomposition, implicit factors, and cycle rounding take
$\softO(m+N)$ time. Since
$\Lambda=8\phi^{-2}=\operatorname{polylog}n$, the coloring
routine determines the total cost:
\begin{align*}
 T_{\rm phase}
 &\le\softO(m+N)
       +\softO\bigl((m+N)(\Lambda/r)^{27/2}\bigr)\\
 &=\softO\bigl((m+N)r^{-27/2}\bigr).
\end{align*}
Run the decomposition, factor construction, and coloring with
failure probability $\zeta/3$ each. If $F_j$ is the event that
the $j$th of these three stages first violates its guarantee,
then
\begin{align*}
 \Pr[\text{phase is unsuccessful}]
 &=\sum_{j=1}^3\Pr[F_j]\\
 &\le3(\zeta/3)=\zeta.
\end{align*}
\end{proof}

\subsection{Dense expander cores and their Gaussian covariance}

The dense phases admit a simpler rounding step. We first remove the
low-degree portions of the expander pieces. The remaining cores have
small covariance even after we regard each matrix as a single vector.
This second covariance bound lets us control the expected Gaussian
matrix norm without imposing a spectral constraint during projection.

Throughout this subsection, let $\ell=\log(2n)$ and assume
$J\le C_J\ell$, as the fixed-grid invariant guarantees. We call a
phase dense when
\begin{equation}\label{eq:dense-regime}
 \rho\le c_{\rm dense}\ell^{-40},
\end{equation}
where $c_{\rm dense}>0$ is a sufficiently small absolute constant.
We keep the parameters $\rho,J,\phi$ of \cref{lem:joint-pieces}.

\begin{lemma}[Dense expander cores]\label{lem:dense-cores}
Suppose \eqref{eq:dense-regime} holds. Set
\begin{equation}\label{eq:dense-min-degree}
 d_0=\left\lfloor c_d\frac{\phi m}{NJ}\right\rfloor
\end{equation}
and $\phi_c=c_{\rm core}\phi$,
for sufficiently small absolute constants $c_d,c_{\rm core}>0$.
From the pieces in \cref{lem:joint-pieces}, we can compute
edge-disjoint cores $\mathcal H_c$ in $\softO(m+N)$ time such that
$\sum_{H\in\mathcal H_c}|E(H)|\ge m/8$ and
$\sum_{H\in\mathcal H_c}|V(H)|\le NJ$, and such that every core has
minimum unweighted degree at least $d_0$ and conductance
$\Phi(H)\ge\phi_c$.
Each core has weight ratio at most two and satisfies
\eqref{eq:erdiam} with the original ambient graph and parameter
$\rho$. Moreover,
\begin{equation}\label{eq:dense-degree-lower}
 d_0\ge c\frac{\phi}{\rho\ell}
       \ge c'\rho^{-1}\ell^{-4}\ge1.
\end{equation}
The running time includes maintaining the active degrees and listing
all removed edges.
\end{lemma}
\begin{proof}
We use the unweighted expander-pruning algorithm of
Saranurak and Wang~\cite[Theorem~1.3]{sw}. We state the consequences
we need with absolute constants. Starting from a graph of conductance
at least $\alpha$, the algorithm processes an adaptive sequence of
at most $\alpha |E|/C_p$ edge deletions. After $t$ deletions, it
removes a set of vertices whose original volume is at most
$C_pt/\alpha$, and the remaining graph has conductance at least
$c_p\alpha$. Its total update time is
$\softO(t/\alpha^2)$, in addition to initialization and reporting
removed edges. The theorem states its conductance guarantee for a
residual graph with degree-restoring self-loops. Removing these loops
leaves every cut boundary unchanged and decreases the volumes in the
denominator. We therefore obtain the same conductance lower bound for
the actual residual graph.

Fix an initial piece $H_0$ and write
$m_H=|E(H_0)|$ and $N_H=|V(H_0)|$. Since its positive weights
differ by at most two, its unweighted support has conductance at
least $\phi/2$. Choose an absolute constant $C_1\ge2C_p$.
If
\begin{equation}\label{eq:dense-discard-test}
 m_H\le C_1d_0N_H/\phi,
\end{equation}
we discard $H_0$. Otherwise we run expander pruning on its unweighted
support. Whenever an active vertex has fewer than $d_0$ active
incident edges, we delete those edges and feed the deletions to the
pruning algorithm. We remove the vertices that pruning reports and
update the active degrees. We also discard isolated vertices, which
contribute no active edges.

A vertex triggers external deletions at most once. Since $H_0$
failed the test \eqref{eq:dense-discard-test}, the number $t_H$ of
external deletions satisfies
\[
 t_H\le d_0N_H<\frac{\phi m_H}{C_1}
       \le\frac{(\phi/2)m_H}{C_p}.
\]
This is within the number of deletions that pruning allows, so all
calls stay within the pruning guarantee. At termination,
every remaining vertex has unweighted degree at least $d_0$.
Pruning guarantees unweighted conductance at least $c_p\phi/2$.
Restoring the original weights gives
\begin{align*}
 \Phi(H)
 &\ge\frac{\min_{e\in H_0}w_e}{\max_{e\in H_0}w_e}
                    \Phi_{\rm unweighted}(H)\\
 &\ge\tfrac12(c_p\phi/2)\ge\phi_c,
\end{align*}
where we choose $c_{\rm core}\le c_p/4$.

The external deletions remove $t_H$ edges. Every other removed edge
meets a pruned vertex, so the original-volume bound gives
\begin{align*}
 |E(H_0)\setminus E(H)|
 &\le t_H+\frac{2C_pt_H}{\phi}\\
 &\le C_2\frac{d_0N_H}{\phi}.
\end{align*}
Isolated vertices that we discard have no remaining incident edges,
so they require no additional edge charge. The same bound, with a
larger $C_2$, covers the pieces discarded in
\eqref{eq:dense-discard-test}. Summing over all initial pieces gives
\begin{align*}
 \sum_{H_0}|E(H_0)\setminus E(H)|
 &\le\frac{C_2d_0}{\phi}\sum_{H_0}|V(H_0)|\\
 &\le\frac{C_2d_0NJ}{\phi}
 \le C_2c_dm\le m/8.
\end{align*}
Choose $c_d\le1/(8C_2)$ and combine this calculation with the initial
coverage $m/4$. This proves the coverage bounds.
Each core uses a subset of its initial vertices and edges, so neither
vertex overlap nor weight ratio increases. The resistance condition
also persists:
\begin{align*}
 \max_{e\in H}w_e\,
     \max_{u,v\in V(H)}\operatorname{ER}_G(u,v)
 &\le\max_{e\in H_0}w_e\,
     \max_{u,v\in V(H_0)}\operatorname{ER}_G(u,v)\\
 &\le\rho.
\end{align*}
We keep only nonempty cores. Their positive conductance makes them
connected.

To account for the floor in \eqref{eq:dense-min-degree}, substitute
$m=16N\log(N+1)/\rho$ and use $J\le C_J\ell$:
\begin{align*}
 c_d\frac{\phi m}{NJ}
 &=\frac{16c_d\phi\log(N+1)}{\rho J}
 \ge c_3\frac{\phi}{\rho\ell}
 =c_4\rho^{-1}\ell^{-4}\\
 &\ge\frac{c_4}{c_{\rm dense}}\ell^{36}\ge2.
\end{align*}
We ensure the last inequality by choosing $c_{\rm dense}$ small
enough. The floor then loses at most a factor of two, proving
\eqref{eq:dense-degree-lower}.

Finally, each edge enters a degree update only when it disappears
from the active graph. The degree queues and edge lists therefore
cost $O(m+NJ)$ operations. The pruning updates cost
\begin{align*}
 \sum_{H_0}\softO(t_H/\phi^2)
 &\le\softO(d_0NJ/\phi^2)
 \le\softO(m/\phi)
 =\softO(m).
\end{align*}
Initialization costs $\softO(m+NJ)$, and the dense regime implies
$NJ=O(m)$. These estimates prove the claimed total time.
\end{proof}

For the cores, use the degree-weighted centers and edge matrices of
\cref{sec:envelope}, with each
center computed from its final core. Let $F$ be their union, let
$k=|F|$, and let $S\subseteq\R^F$ impose the degree equations
separately in each core. Write $\Pi$ for the orthogonal projector
onto $S$. The next lemma concerns both the exact matrix map
$\mathcal X$ and its implicit counterpart
$\overline{\mathcal X}$ from \cref{lem:implicit-factors}.

\begin{lemma}[Gaussian norm on the dense cores]
\label{lem:dense-gaussian}
Let $g\sim N(0,I_k)$, and define
$G=\mathcal X(\Pi g)$ and
$\bar G=\overline{\mathcal X}(\Pi g)$. Conditional on the Gram bounds
of \cref{lem:implicit-factors}, the matrix variances satisfy
$\|\mathbb E G^2\|\le\rho$ and $\|\mathbb E\bar G^2\|\le2\rho$, and the
vectorized covariance of $G$ satisfies
\begin{equation}\label{eq:dense-vector-variance}
 \left\|\operatorname{Cov}(\operatorname{vec}G)\right\|
 \le\frac{C}{d_0^2\phi_c^4}.
\end{equation}
The same bound, with a larger constant $C'$, holds for $\bar G$.
Here $\operatorname{vec}$ stacks the columns of a matrix. Each
$Y\in\{G,\bar G\}$ satisfies
\begin{equation}\label{eq:dense-gaussian-norm}
 \mathbb E\|Y\|
 \le C\left(\sqrt\rho+\rho\ell^{23/2}\right)
 \le C'\sqrt\rho.
\end{equation}
The implicit matrices have physical dimension $2m$, but their common
image has dimension at most $N$. We apply their linear combinations
in $\softO(m+N)$ time and never form their vectorized covariance.
\end{lemma}
\begin{proof}
Abbreviate $R_L=R_L(A)$ and $R_R=R_R(A)$. For a core $H$, let
$w_H=\max_{e\in H}w_e$, $d_L=A_H\1$, and
$s_H=\1^\top d_L$. Define the centering matrix
$T_H^L$ to have columns $e_u-\mu_L^H$ for $u$ on its left side.
On these coordinates,
$T_H^L\operatorname{diag}(d_L)(T_H^L)^\top
=\operatorname{diag}(d_L)-d_Ld_L^\top/s_H$. The minimum unweighted
degree and the weight ratio at most two give
$\operatorname{diag}(d_L)\succeq\tfrac12d_0w_HI$. Set
$U_H=w_H\sum_{u\in V_L(H)}a_u^H(a_u^H)^\top$ and
$V_H=w_H\sum_{v\in V_R(H)}b_v^H(b_v^H)^\top$.
The preceding identities and the Schur gap
\eqref{eq:expander-defect}, now with conductance $\phi_c$, give
\begin{align*}
 U_H
 &=w_H R_L^{\dagger/2}T_H^L(T_H^L)^\top R_L^{\dagger/2}\\
 &\preceq\frac{2}{d_0}R_L^{\dagger/2}
       \left(\operatorname{diag}(d_L)-\frac{d_Ld_L^\top}{s_H}\right)
                                           R_L^{\dagger/2}\\
 &\preceq\frac{4}{d_0\phi_c^2}
                R_L^{\dagger/2}R_L(A_H)R_L^{\dagger/2}.
\end{align*}
Summing over the cores and using \eqref{eq:superadd} yields
\begin{align}
 \sum_HU_H
 &\preceq\frac{4}{d_0\phi_c^2}R_L^{\dagger/2}
                   \left(\sum_HR_L(A_H)\right)R_L^{\dagger/2}\notag\\
 &\preceq\frac{4}{d_0\phi_c^2}I.
                                           \label{eq:dense-side-total}
\end{align}
Applying the same calculation to $A^\top$ gives
$\sum_HV_H\preceq4d_0^{-1}\phi_c^{-2}I$.

For $e=(u,v)\in H$, write
$C_e=w_ea_u^H(b_v^H)^\top$, so that $X_e$ is its symmetric dilation.
The input has at most one arc per ordered pair, and all subsequent
operations only change weights or delete arcs. Thus each core contains
at most one edge for each pair $(u,v)$. Since every summand below is
positive semidefinite,
\begin{align*}
 \sum_{e\in E(H)}\operatorname{vec}(C_e)
                         \operatorname{vec}(C_e)^\top
 &=\sum_{(u,v)\in E(H)}w_e^2
           \bigl(b_v^H(b_v^H)^\top\bigr)
              \otimes\bigl(a_u^H(a_u^H)^\top\bigr)\\
 &\preceq w_H^2\sum_{u\in V_L(H),\,v\in V_R(H)}
           \bigl(b_v^H(b_v^H)^\top\bigr)
              \otimes\bigl(a_u^H(a_u^H)^\top\bigr)\\
 &=V_H\otimes U_H.
\end{align*}
Apply \eqref{eq:dense-side-total} and its right-side analogue to obtain
\begin{align}
 \sum_{e\in F}\operatorname{vec}(C_e)
                         \operatorname{vec}(C_e)^\top
 &\preceq\sum_HV_H\otimes U_H\notag\\
 &\preceq\left(\sum_HV_H\right)\otimes
                         \left(\sum_HU_H\right)\notag\\
 &\preceq\frac{16}{d_0^2\phi_c^4}I.
                                           \label{eq:dense-rect-cov}
\end{align}
Let $K_C$ have columns $\operatorname{vec}(C_e)$. Projection changes
the rectangular covariance to $K_C\Pi K_C^\top$, so
\begin{align*}
 \operatorname{Cov}\!\left(
       \operatorname{vec}\sum_e(\Pi g)_eC_e\right)
 &=K_C\Pi K_C^\top\preceq K_CK_C^\top
 \preceq\frac{16}{d_0^2\phi_c^4}I.
\end{align*}
Symmetric dilation multiplies squared Frobenius norms by two.
Its vectorization map therefore has squared operator norm two,
which proves \eqref{eq:dense-vector-variance}.
The projected-variance proof in \cref{lem:fast-envelope} applies to
the cores without change, because they keep the same resistance
parameter. It gives
$\mathbb E G^2=\sum_{e\in F}\mathcal X(\Pi e_e)^2\preceq\rho I$.

For the implicit factors, the proof of \cref{lem:implicit-factors}
provides a fixed map $U$ with $\|U\|\le1+\xi$ and
$\bar X_e=UX_eU^\top$. Hence $\bar G=UGU^\top$, and
\begin{align*}
 \left\|\operatorname{Cov}(\operatorname{vec}\bar G)\right\|
 &\le\|U\otimes U\|^2
          \left\|\operatorname{Cov}(\operatorname{vec}G)\right\|\\
 &\le\frac{C(1+\xi)^4}{d_0^2\phi_c^4}.
\end{align*}
Similarly,
$\mathbb E\bar G^2\preceq\|U\|^2U(\mathbb E G^2)U^\top
\preceq(1+\xi)^4\rho I\preceq2\rho I$.
This proves both covariance assertions.

We now apply the Gaussian matrix norm estimate of Bandeira, Boedihardjo,
and van Handel~\cite[Corollary~2.2 and Lemma~2.5]{bbvh}.
For a centered self-adjoint Gaussian matrix
$Y$ acting on a $d$-dimensional space, define
$\sigma(Y)^2=\|\mathbb E Y^2\|$ and
$v(Y)^2=\|\operatorname{Cov}(\operatorname{vec}Y)\|$.
Their estimate, together with the bound for the associated free
Gaussian model, gives
\begin{equation}\label{eq:dense-bbvh}
 \mathbb E\|Y\|
 \le C\left(\sigma(Y)
          +\sqrt{\sigma(Y)v(Y)}\,\log^{3/4}(2d)\right).
\end{equation}
All exact matrices act on the two defect images, whose combined
dimension is at most $N$. All implicit matrices act on
$\operatorname{im}Z_L\oplus\operatorname{im}Z_R$, also of dimension
at most $N$. We restrict to these common images when applying
\eqref{eq:dense-bbvh}. Orthogonal compression preserves the matrix
norm and cannot increase the vectorized covariance norm.

For either $Y=G$ or $Y=\bar G$, the established bounds give
$\sigma(Y)\le C\sqrt\rho$ and
$v(Y)\le C/(d_0\phi_c^2)\le C\rho\ell^{10}$.
The last inequality uses \eqref{eq:dense-degree-lower} and
$\phi_c=\Theta(\ell^{-3})$. Since $N\le2n$, Young's inequality in
\eqref{eq:dense-bbvh} now yields
\begin{align*}
 \mathbb E\|Y\|
 &\le C\left(\sqrt\rho
        +\sqrt{\sqrt\rho\,v(Y)}\,\ell^{3/4}\right)\\
 &\le C'\left(\sqrt\rho+v(Y)\ell^{3/2}\right)\\
 &\le C''\left(\sqrt\rho+\rho\ell^{23/2}\right).
\end{align*}
Finally, \eqref{eq:dense-regime} gives
$\rho\ell^{23/2}=\sqrt\rho\,\bigl(\sqrt\rho\,\ell^{23/2}\bigr)
\le\sqrt{c_{\rm dense}}\,\ell^{-17/2}\sqrt\rho\le C\sqrt\rho$.
This proves \eqref{eq:dense-gaussian-norm}.

The implementation uses the final-core centers in the aggregation
formulas of \cref{lem:implicit-factors}. Their cost is
$O(|F|+\sum_H|V(H)|)=O(m+NJ)=O(m)$, followed by the same implicit
Laplacian operations. The PSD envelope becomes
$\sum_{e\in F}\bar D_e\preceq\Lambda_c I$ with $\Lambda_c=8\phi_c^{-2}$.
Thus the physical dimension $2m$ incurs only linear storage and
nearly linear matrix-vector costs. The covariance matrices above
serve only to prove the Gaussian bound.
\end{proof}

\subsection{Gaussian projection onto the degree-constrained cube}
\label{sec:dense-cube-projection}

For the dense phases, we project a Gaussian vector onto the cube and
the degree constraints alone. The following comparison transfers the
Gaussian matrix-norm bound to this projection. It lets us remove the
spectral constraint from the optimization problem. The proof uses
the Gaussian rotation argument of Maurey and Pisier~\cite{pisier}.

\begin{lemma}[A Gaussian comparison for cube projection]
\label{lem:cube-comparison}
Let $S\subseteq\R^q$ be a linear subspace with orthogonal projector
$\Pi_S$, and let $P$ denote Euclidean projection onto
$S\cap[-1,1]^q$. For any linear map $\mathcal T$ from $\R^q$ to a
finite-dimensional normed space and $G\sim\mathcal N(0,I_q)$,
\begin{equation}
 \mathbb E\|\mathcal T(PG)\|
 \le\frac\pi2\,\mathbb E\|\mathcal T(\Pi_SG)\|.
 \label{eq:cube-gaussian-comparison}
\end{equation}
\end{lemma}
\begin{proof}
Projection onto a closed convex set is $1$-Lipschitz. The set
$S\cap[-1,1]^q$ is symmetric, so $P$ is odd and $\mathbb E P(G)=0$.
For independent standard Gaussians $G,H$, Jensen's inequality gives
\begin{align}
 \mathbb E\|\mathcal T(PG)\|
 &=\mathbb E_G\left\|\mathbb E_H
                  \mathcal T(PG-PH)\right\|\notag\\
 &\le\mathbb E\|\mathcal T(PG-PH)\|.
 \label{eq:cube-rotation-jensen}
\end{align}
For $0\le t\le\pi/2$, define $G_t=G\cos t+H\sin t$ and
$H_t=-G\sin t+H\cos t$.
For each fixed $t$, the vectors $G_t,H_t$ are independent standard
Gaussians. Rademacher's theorem gives a derivative $DP$ almost
everywhere. The map $t\mapsto P(G_t)$ is absolutely continuous.
Since $G_t$ has a density for each $t$, Fubini's theorem shows that
the chain rule applies for almost every $t$ with probability one.
Consequently, $PG-PH=-\int_0^{\pi/2}DP(G_t)H_t\,dt$, and therefore
\begin{equation}
 \mathbb E\|\mathcal T(PG-PH)\|
 \le\int_0^{\pi/2}
       \mathbb E\|\mathcal T(DP(G_t)H_t)\|\,dt.
 \label{eq:cube-rotation-integral}
\end{equation}
Where $J=DP(G_t)$ exists, its operator norm is at most one and its
image lies in $S$. Thus every $u\in\R^q$ satisfies
$u^\top JJ^\top u=\|J^\top\Pi_Su\|_2^2\le\|\Pi_Su\|_2^2$, which means
$JJ^\top\preceq\Pi_S$.
Condition on $G_t$. The vector $JH_t$ is Gaussian with covariance
$JJ^\top$. Add an independent centered Gaussian $Z$ with covariance
$\Pi_S-JJ^\top$. Jensen's inequality and covariance addition give
\begin{align*}
 \mathbb E_{H_t}\|\mathcal T(JH_t)\|
 &\le\mathbb E_{H_t,Z}\|\mathcal T(JH_t+Z)\|\\
 &=\mathbb E\|\mathcal T(\Pi_SG)\|.
\end{align*}
Substitute this bound into \eqref{eq:cube-rotation-integral}, and
then use \eqref{eq:cube-rotation-jensen}. The integration interval
has length $\pi/2$, which proves the claim.
\end{proof}

We next give an explicit implementation of the cube projection on the
dense expander cores. Its dual variables are vertex potentials.
Each iteration clips the edge coordinates and adjusts each potential
by its current degree imbalance. A random set of edges stays strictly
inside the cube throughout the iteration. These edges keep enough
expansion to make the iteration converge geometrically.

\begin{lemma}[Fast Gaussian cube projection on dense expanders]
\label{lem:dense-cube}
Let $\mathcal H$ be a nonempty family of connected, edge-disjoint
bipartite graphs with a total of $q\le m$ edges, where $m\ge2$.
Make a separate copy of each vertex in each graph. Assume that every
graph has conductance at least $\phi\in(0,1]$, edge-weight ratio at
most two, and minimum combinatorial degree at least $d$, where
\begin{equation}
 d\ge C\phi^{-4}\log^3(2m)
 \label{eq:dense-cube-degree}
\end{equation}
for a sufficiently large absolute constant $C$. Let $S\subseteq\R^q$
impose the exact degree constraints within each graph, and draw
$g\sim\mathcal N(0,I_q)$. Let
$p_0=\Pr[|Z|\ge3/2]$ for $Z\sim\mathcal N(0,1)$, set
$\gamma=p_0/4$, and let $x_*=\operatorname{proj}_{S\cap[-1,1]^q}(g)$.
For every $0<\tau<1$, a fixed number of arithmetic operations,
bounded by
\begin{equation}
 O\!\left(q\phi^{-2}\log\frac{Cm^2}{\tau}\right),
 \label{eq:dense-cube-time}
\end{equation}
suffices to return $y\in S\cap[-1,1]^q$. With probability at least
$0.98$, the returned vector also satisfies $\|y-x_*\|_2\le\tau$ and
$|\{e:|y_e|\ge1-\tau\}|\ge2\gamma q$.
On this event, one of $y,-y$ has at least $\gamma q$ coordinates
at most $-1+\tau$. The algorithm returns a feasible cube point even
when this event fails. Its iteration count and the degree threshold
do not depend on a requested final failure probability.
\end{lemma}
\begin{proof}
We first analyze one graph $H$ with $s$ edges. Divide its weights by
its minimum weight, so $1\le w_e\le2$. This rescaling does not change
its degree-constraint space. Orient the edges from left to right,
let $B$ be the edge-by-vertex incidence matrix, and set
$W=\diag(w)$, $L=B^\top W^2B$, $D=\diag(L)$, and $V_H=\1^\top D\1$.
We use $b_e^\top$ for row $e$ of $B$. The graph with edge weights
$w_e^2$ has conductance at least $\phi/2$. Cheeger's inequality
therefore gives
\begin{equation}
 \lambda_2(D^{-1/2}LD^{-1/2})\ge c\phi^2.
 \label{eq:dense-cube-gap}
\end{equation}
Every vertex has $D_{vv}\ge d$, and $V_H\le8s$.

Define $\operatorname{clip}(z)=\max\{-1,\min\{1,z\}\}$, applied
coordinatewise to vectors. Its convex antiderivative is
\begin{equation*}
 \psi(z)=
 \begin{cases}
 z^2/2,& |z|\le1,\\
 |z|-1/2,& |z|>1.
 \end{cases}
\end{equation*}
The dual objective is
\begin{equation}
 F(p)=\sum_{e\in E(H)}\psi(g_e-w_eb_e^\top p),
 \label{eq:dense-cube-dual}
\end{equation}
and the degree imbalance is $b(p)=B^\top W\operatorname{clip}(g-WBp)$.
They satisfy $\nabla F(p)=-b(p)$. At differentiability points of
this gradient, its Hessian is the active-edge Laplacian
\begin{equation}
 \nabla^2F(p)
 =B^\top W^2
   \diag\bigl(\1_{\{|g_e-w_eb_e^\top p|<1\}}\bigr)B.
 \label{eq:dense-cube-hessian}
\end{equation}

\paragraph{The random interior edges and initial imbalance.}
Let $H_0$ contain the edges with $|g_e|\le1/2$, and let $L_0$ be its
Laplacian with edge weights $w_e^2$. Each edge enters independently
with the same absolute probability $p_{\rm int}>0$.
The gap in \eqref{eq:dense-cube-gap} bounds the leverage of every
edge $e=(u,v)$:
\begin{equation}
 w_e^2b_e^\top L^\dagger b_e
 \le\frac{w_e^2}{c\phi^2}
             \left(\frac1{D_{uu}}+\frac1{D_{vv}}\right)
 \le\frac{C_1}{\phi^2d}.
 \label{eq:dense-cube-leverage}
\end{equation}
Apply the matrix Chernoff bound~\cite[Corollary~5.2 and Remark~5.3]{tropp} to
$w_e^2L^{\dagger/2}b_eb_e^\top L^{\dagger/2}$ on
$\operatorname{im}L$. Their sum is the identity on that image.
For an absolute constant $c_1>0$, this gives
\begin{equation}
 \Pr[L_0\nsucceq(p_{\rm int}/2)L]
 \le |V(H)|\exp(-c_1\phi^2d).
 \label{eq:dense-cube-backbone-failure}
\end{equation}
The summands in $b(0)_v$ are independent, centered, and bounded in
absolute value by their edge weights. With
$a=\sqrt{2\log(1600m)/d}$, Hoeffding's inequality gives
\begin{equation}
 \Pr[|b(0)_v|>aD_{vv}]
 \le2\exp(-a^2D_{vv}/2).
 \label{eq:dense-cube-imbalance-failure}
\end{equation}
Let $\mathcal E$ be the event that all graphs satisfy
$L_0\succeq(p_{\rm int}/2)L$ and
$\|D^{-1}b(0)\|_\infty\le a$. There are at most $2q\le2m$
vertex copies. Choosing $C$ in \eqref{eq:dense-cube-degree}
sufficiently large gives the explicit failure estimate
\begin{align}
 \Pr[\mathcal E^c]
 &\le\sum_{H\in\mathcal H}|V(H)|e^{-c_1\phi^2d}
       +\sum_{H\in\mathcal H}\sum_{v\in V(H)}
                         2e^{-a^2D_{vv}/2}\notag\\
 &\le2m e^{-c_1\phi^2d}
       +4m e^{-\log(1600m)}
 \le\frac1{400}+\frac1{400}=\frac1{200}.
 \label{eq:dense-cube-good-probability}
\end{align}
We work on $\mathcal E$ until we account for the remaining failure
probability.

\paragraph{A uniform bound for inverse Laplacians.}
Fix a Laplacian $A$ with $L_0\preceq A\preceq L$ whose edge
weights lie between zero and the corresponding weights of $L$.
Set $Q_A=I-\tfrac12D^{-1}A$ and $f=D^{-1}b(0)$.
The matrix $Q_A$ is a nonnegative lazy transition matrix with
stationary distribution $\pi_v=D_{vv}/V_H$. Its spectral gap is at
least $c_2\phi^2$, and $\pi_{\min}\ge1/(8m)$. Also
$\sum_v\pi_vf_v=0$ and $\|f\|_\infty\le a$. Hence
\begin{align}
 \|Q_A^tf\|_\infty
 &\le\min\{\|f\|_\infty,
             \pi_{\min}^{-1/2}\|Q_A^tf\|_{2,\pi}\}\notag\\
 &\le a\min\{1,\sqrt{8m}\,e^{-c_2\phi^2t}\}.
 \label{eq:dense-cube-inverse-decay}
\end{align}
The vector $u=\tfrac12\sum_{t\ge0}Q_A^tf$ satisfies $Au=b(0)$
and $\1^\top Du=0$. Splitting this series at
$t_0=\lceil C_2\phi^{-2}\log(2m)\rceil$ gives
\begin{align}
 \|u\|_\infty
 &\le\frac a2\left(t_0+
          \frac{\sqrt{8m}\,e^{-c_2\phi^2t_0}}
               {1-e^{-c_2\phi^2}}\right)\notag\\
 &\le C_3\phi^{-2}\log(2m)\,a=:R\le\frac1{16}.
 \label{eq:dense-cube-potential-radius}
\end{align}
The last inequality follows from \eqref{eq:dense-cube-degree}.
The same bound holds for every such $A$.

\paragraph{Existence of a small dual optimum.}
Let $K=\{p:\1^\top Dp=0,\ \|p\|_\infty\le2R\}$.
For $p\in K$, every edge shift has magnitude at most $8R\le1/2$.
Define
\begin{equation}
 a_e(p)=
 \frac{\operatorname{clip}(g_e)
       -\operatorname{clip}(g_e-w_eb_e^\top p)}{w_eb_e^\top p}
 \label{eq:dense-cube-averaged-hessian}
\end{equation}
and the averaged Hessian $\bar L(p)=B^\top W^2\diag(a_e(p))B$.
At a zero denominator, define $a_e(p)$ by the derivative of clipping
at $g_e$. Almost surely $g_e\ne\pm1$, so this extension is
continuous. We have $0\le a_e(p)\le1$. Every edge of $H_0$ has
$a_e(p)=1$, because its entire interpolation segment lies in
$[-1,1]$. Thus $L_0\preceq\bar L(p)\preceq L$. The definition of
$a_e(p)$ also gives $b(p)=b(0)-\bar L(p)p$.
Let $d_H=D\1$. The map
\begin{equation}
 T(p)=\left(\bar L(p)+\frac{d_Hd_H^\top}{V_H}\right)^{-1}b(0)
 \label{eq:dense-cube-brouwer-map}
\end{equation}
is continuous on $K$. The rank-one term makes the matrix invertible.
Multiplying its defining equation by $\1^\top$ shows that
$d_H^\top T(p)=0$. Therefore $T(p)$ is exactly the mean-zero
solution of $\bar L(p)T(p)=b(0)$ from the preceding paragraph.
In particular, $\|T(p)\|_\infty\le R$, so $T$ maps $K$ into itself.
Brouwer's fixed-point theorem gives a point $p_*=T(p_*)$, which
satisfies $\|p_*\|_\infty\le R$ and $b(p_*)=b(0)-\bar L(p_*)p_*=0$.
The vector $x_*^H=\operatorname{clip}(g-WBp_*)$ is the Euclidean
projection onto $\ker(B^\top W)\cap[-1,1]^{E(H)}$. To see this
directly, it minimizes
$\tfrac12\|x-g\|_2^2+p_*^\top B^\top Wx$ over the cube and
satisfies $B^\top Wx_*^H=0$. Every feasible competitor has zero
linear term, so $x_*^H$ minimizes the original squared distance.

\paragraph{Geometric convergence by vertex updates.}
Starting at $p^{(0)}=0$, perform
\begin{equation}
 p^{(t+1)}=p^{(t)}+\tfrac12D^{-1}b(p^{(t)}).
 \label{eq:dense-cube-jacobi}
\end{equation}
Each update uses $O(s)$ operations. Write $e^{(t)}=p^{(t)}-p_*$.
As long as $p^{(t)}\in K$, integrate the active-edge Hessians along
the segment from $p_*$ to $p^{(t)}$ to obtain a Laplacian $A_t$ with
$L_0\preceq A_t\preceq L$ and
\begin{equation}
 e^{(t+1)}=(I-\tfrac12D^{-1}A_t)e^{(t)}.
 \label{eq:dense-cube-error-update}
\end{equation}
The update matrix is nonnegative and row-stochastic. It preserves
$D$-weighted mean zero. Therefore
$\|e^{(t+1)}\|_\infty\le\|e^{(t)}\|_\infty\le R$, and
$\|p^{(t+1)}\|_\infty\le\|p_*\|_\infty+\|e^{(t+1)}\|_\infty\le2R$.
This proves $p^{(t)}\in K$ for every $t$ by induction. On the
$D$-mean-zero subspace, the eigenvalues of
$D^{-1/2}A_tD^{-1/2}$ lie between $c_3\phi^2$ and $2$.
Consequently,
$\|e^{(t)}\|_D\le(1-c_4\phi^2)^t\|p_*\|_D\le\sqrt{8s}\,e^{-c_4\phi^2t}$.
Since clipping is $1$-Lipschitz,
\begin{align}
 \|\operatorname{clip}(g-WBp^{(t)})-x_*^H\|_2
 &\le\|WBe^{(t)}\|_2\notag\\
 &\le\sqrt2\|e^{(t)}\|_D
 \le4\sqrt s\,e^{-c_4\phi^2t}.
 \label{eq:dense-cube-primal-convergence}
\end{align}
For a prescribed $\delta_H>0$, fix in advance
$T_H=\lceil C_4\phi^{-2}\log(8\sqrt s/\delta_H)\rceil$ iterations
and let $\hat x=\operatorname{clip}(g-WBp^{(T_H)})$.
On $\mathcal E$, this gives $\|\hat x-x_*^H\|_2\le\delta_H$.
We use the same iteration cap for every Gaussian input.

\paragraph{Exact degree correction.}
Route the residual $b=B^\top W\hat x$ on a spanning tree of $H$.
If $f$ denotes the resulting signed edge flow, set $h=W^{-1}f$.
Then $B^\top Wh=b$. Each tree edge carries a signed sum of a
subset of the demands, so on $\mathcal E$,
\begin{align}
 \|h\|_\infty\le\|b\|_1
 &=\|B^\top W(\hat x-x_*^H)\|_1\notag\\
 &\le2\sum_e w_e|\hat x_e-x_{*,e}^H|
 \le4\sqrt s\,\delta_H.
 \label{eq:dense-cube-tree-correction}
\end{align}
Since $h$ is supported on at most $s$ tree edges, also
$\|h\|_2\le\sqrt s\,\|h\|_\infty\le4s\delta_H$.
Return
\begin{equation}
 y^H=\frac{\hat x-h}{1+\|h\|_\infty}.
 \label{eq:dense-cube-feasible-output}
\end{equation}
For every input $g$, this vector belongs to
$\ker(B^\top W)\cap[-1,1]^{E(H)}$. On $\mathcal E$, its error is
\begin{align}
 \|y^H-x_*^H\|_2
 &\le\|\hat x-h-x_*^H\|_2
       +\frac{\|h\|_\infty}{1+\|h\|_\infty}
                          \|\hat x-h\|_2\notag\\
 &\le\delta_H+\|h\|_2+\sqrt s\|h\|_\infty
 \le(1+8s)\delta_H.
 \label{eq:dense-cube-repaired-error}
\end{align}
Let $h_{\rm pieces}=|\mathcal H|$ and choose
$\delta_H=\tau/\bigl(\sqrt{h_{\rm pieces}}(1+8|E(H)|)\bigr)$.
The global feasible set is the product of the feasible sets of the
pieces. Concatenating their outputs therefore gives
\begin{align*}
 \|y-x_*\|_2^2
 &=\sum_H\|y^H-x_*^H\|_2^2
 \le\sum_H\frac{\tau^2}{h_{\rm pieces}}=\tau^2.
\end{align*}
The prescribed iteration caps cost
\begin{align*}
 \sum_H O(|E(H)|T_H)
 &\le O\!\left(q\phi^{-2}\log\frac{Cm^2}{\tau}\right).
\end{align*}
The spanning-tree corrections take $O(q)$ additional time.

\paragraph{Many coordinates reach the boundary.}
On $\mathcal E$, the dual bound gives $|w_eb_e^\top p_*|\le4R\le1/4$.
Hence every edge with $|g_e|\ge3/2$ has
$|g_e-w_eb_e^\top p_*|\ge5/4$, and therefore $|x_{*,e}|=1$.
The count $Z=|\{e:|g_e|\ge3/2\}|$ is binomial with mean $p_0q$, and
the Chernoff bound gives $\Pr[Z<p_0q/2]\le e^{-p_0q/8}$.
The degree assumption, with a sufficiently large absolute constant,
ensures $q\ge d$ and $e^{-p_0q/8}\le1/100$. Thus the probability
that either the projection analysis or this count fails is at most
\begin{align*}
 \Pr[\mathcal E^c]+\Pr[Z<p_0q/2]
 &\le\frac1{200}+\frac1{100}
 <0.02.
\end{align*}
On the complementary event, every one of these $Z$ coordinates
satisfies $|y_e|\ge1-\tau$, because
$\|y-x_*\|_\infty\le\|y-x_*\|_2\le\tau$. Therefore
$|\{e:|y_e|\ge1-\tau\}|\ge p_0q/2=2\gamma q$, and one of $y,-y$ has at
least $\gamma q$ coordinates at most $-1+\tau$.
This proves the approximation and boundary-count guarantees.
The explicit correction \eqref{eq:dense-cube-feasible-output}
and the fixed iteration caps apply regardless of the random events.
\end{proof}

\begin{algorithm}[H]
\caption{\Call{CubeProject}: Gaussian projection with exact degree repair}
\label{alg:dense-cube}
\small
\textbf{Input:} expander cores $\mathcal H$, Gaussian edge vector $g$,
accuracy $0<\tau<1$, and conductance lower bound $\phi_c$.
\begin{enumerate}[label=\arabic*:,leftmargin=2em,itemsep=3pt]
\item Set $h_{\rm pieces}\gets|\mathcal H|$.
\item \textbf{for each} core $H\in\mathcal H$ \textbf{do}:
\item \hspace*{1em}Normalize its edge weights into $[1,2]$ and form
      $W$, its incidence matrix $B$, and $D=\diag(B^\top W^2B)$.
\item \hspace*{1em}Set $s\gets|E(H)|$,
      $\delta_H\gets\tau/(\sqrt{h_{\rm pieces}}(1+8s))$, and
      $T_H\gets\lceil C\phi_c^{-2}\log(8\sqrt s/\delta_H)\rceil$.
      Initialize $p\gets0$.
\item \hspace*{1em}\textbf{repeat} exactly $T_H$ times:
      $p\gets p+\tfrac12D^{-1}B^\top W
                    \operatorname{clip}(g_H-WBp)$.
\item \hspace*{1em}Set $\hat x\gets\operatorname{clip}(g_H-WBp)$.
      Route $B^\top W\hat x$ on a spanning tree to obtain $f$ with
      $B^\top f=B^\top W\hat x$, and set $h\gets W^{-1}f$.
\item \hspace*{1em}Set $y^H\gets(\hat x-h)/(1+\|h\|_\infty)$.
\item \textbf{end for}. \textbf{Return} the concatenation of the $y^H$.
\end{enumerate}
\end{algorithm}

The comparison in \cref{lem:cube-comparison} controls the discrepancy
of the exact projection. The projected variance controls the cost of
approximating it. In particular, if
$\sum_i\mathcal X(\Pi_Se_i)^2\preceq\rho I$, then for $z\in S$,
\begin{align}
 \|\mathcal X(z)\|
 &=\sup_{\|u\|_2=1}
       \left\|\sum_i z_i\mathcal X(\Pi_Se_i)u\right\|_2\notag\\
 &\le\|z\|_2\sup_{\|u\|_2=1}
       \left(\sum_i\|\mathcal X(\Pi_Se_i)u\|_2^2\right)^{1/2}
 \le\sqrt\rho\,\|z\|_2.
 \label{eq:dense-cube-variance-lipschitz}
\end{align}
Taking $\tau$ below the desired near-boundary tolerance therefore
preserves both the discrepancy and the number of small weights.
The preceding algorithm needs only logarithmically many additional
iterations as we decrease this tolerance.

\subsection{A phase with nearly linear running time}

We now combine the two projection routines. The spectral-projection
routine handles the range where its inverse-radius cost is
polylogarithmic. The dense routine uses ordinary cube projection and
checks its discrepancy afterwards. The check needs only a constant
factor estimate of the norm.
We cap every promise-based subroutine at its stated operation count
and report failure if it exceeds that count. This includes pruning
after a randomized decomposition. On the events where the input
promises hold, the proved bounds prevent a timeout. These caps ensure
the running-time bound even on unsuccessful executions.

\begin{lemma}[Checking a norm by implicit products]
\label{lem:implicit-norm-check}
Let $M$ be a real symmetric $d\times d$ matrix available through
matrix-vector products, and let $0<\delta<1/2$. In
$O(\log(2d/\delta))$ products, we can compute $U\ge0$ such that
$U\le2\|M\|$ always and $\Pr[\|M\|\le U]\ge1-\delta$.
The product count does not depend on $\|M\|$ or its spectral gap.
\end{lemma}
\begin{proof}
Draw $h\sim\mathcal N(0,I_d)$ and set
$z=(M^2)^kh$, where $k=C\lceil\log(2d/\delta)\rceil$.
Return $U=2\|Mz\|_2/\|z\|_2$ if $z\ne0$, and $U=0$ otherwise.
The upper bound $U\le2\|M\|$ follows directly. If $M=0$, both
claims hold. Otherwise, let $\lambda=\|M\|^2$ and choose a unit
top eigenvector $u$ of $M^2$. Gaussian small-ball and Markov bounds
give $\Pr[|u^\top h|<\delta/4]\le\delta/2$ and
$\Pr[\|h\|_2^2>2d/\delta]\le\delta/2$. Hence the event that
$|u^\top h|\ge\delta/4$ and $\|h\|_2^2\le2d/\delta$ has probability
at least $1-\delta$.
On this event, decompose $z=z_{\rm hi}+z_{\rm lo}$ into the
eigenspaces of $M^2$ with eigenvalues at least $\lambda/2$ and below
$\lambda/2$. Then $\|z_{\rm hi}\|_2^2\ge\lambda^{2k}(u^\top h)^2$ and
$\|z_{\rm lo}\|_2^2\le(\lambda/2)^{2k}\|h\|_2^2$, so
\[
 \frac{\|z_{\rm lo}\|_2^2}{\|z_{\rm hi}\|_2^2}
 \le2^{-2k}\frac{\|h\|_2^2}{(u^\top h)^2}
 \le 2^{-2k}\frac{32d}{\delta^3}\le1.
\]
Consequently,
\[
 \frac{\|Mz\|_2^2}{\|z\|_2^2}
 \ge\frac\lambda2
       \frac{\|z_{\rm hi}\|_2^2}
            {\|z_{\rm hi}\|_2^2+\|z_{\rm lo}\|_2^2}
 \ge\frac\lambda4.
\]
Thus $U\ge\sqrt\lambda=\|M\|$ on the stated event.
\end{proof}

The dense-core lemmas use a sufficiently small absolute constant
$c_{\rm dense}>0$ in the threshold
$\rho\le c_{\rm dense}\log^{-40}(2n)$.
We enlarge the constant $C_{\rm pc}$ in the phase radius to cover
both branches. All constants below are absolute. In a dense phase,
write $\phi_c$ for the conductance lower bound of the cores,
so that $\phi_c=\Theta(\phi)$, and set $\Lambda_c=8\phi_c^{-2}$.

\begin{algorithm}[H]
\caption{\Call{ImprovedPhase}: selecting the projection routine}
\label{alg:hybrid-phase}
\small
\textbf{Input:} current adjacency $A$, radius $r$, failure probability $\zeta$.
\begin{enumerate}[label=\arabic*:,leftmargin=2em,itemsep=3pt]
\item Compute $m,N,W,J,\rho$ as in \cref{lem:joint-pieces}.
      Set $\ell\gets\log(2n)$.
\item If $\rho>c_{\rm dense}\ell^{-40}$, \textbf{return}
      $\Call{SpectralPhase}(A,r,\zeta)$ using \cref{alg:fast-phase}.
\item Find the cores of \cref{lem:dense-cores}, with failure
      probability $\zeta/4$. Set $F\gets\bigcup_HE(H)$ and $q\gets|F|$.
\item Build the implicit factors of \cref{lem:implicit-factors},
      with failure probability $\zeta/4$, using the cores' centers.
      Set $\beta\gets r/(8\Lambda_c)$ and $\tau\gets\beta/4$.
\item Set $R\gets\lceil C\log(4/\zeta)\rceil$ and
      $\delta\gets\zeta/(4R)$.
\item \textbf{repeat} at most $R$ times:
\item \hspace*{1em}Draw $g\sim\mathcal N(0,I_q)$ and run the capped
      cube-projection routine of \cref{alg:dense-cube} to accuracy
      $\tau$, obtaining $y\in S\cap[-1,1]^F$.
\item \hspace*{1em}Choose the sign of $y$ to maximize
      $|\{e:y_e\le-1+\beta\}|$.
      If this count is below $\gamma_c q$, continue.
\item \hspace*{1em}Apply \cref{lem:implicit-norm-check} to
      $\overline{\mathcal X}(y)$ with failure probability $\delta$.
      If its output satisfies $U\le r/2$, accept $y$ and exit.
\item \textbf{end repeat}. If no trial passes both checks,
      \textbf{return failure}.
\item Replace $w_e$ by $w_e(1+y_e)$ on $F$. Apply the cycle
      rounding of \cref{lem:fast-cycle} on the coordinates near $-1$.
      Keep all other edges fixed, delete zero weights, and
      \textbf{return} the resulting adjacency $A'$.
\end{enumerate}
\end{algorithm}

For clarity, the ordinary projection in the dense branch minimizes
Euclidean distance subject only to the cube and degree equations.
It does not optimize a spectral constraint. The covariance and
rotation lemmas make the discrepancy check pass with constant
probability. The fixed iteration cap and exact repair ensure that
every trial has bounded running time and returns a feasible cube
point, including trials on which the probabilistic projection
estimate fails.

\begin{lemma}[Nearly linear phase]
\label{lem:hybrid-phase}
Assume $J\le C_J\log(2n)$, where $C_J$ is the absolute constant
provided by grid rounding. There are absolute constants
$C_0,C_{\rm pc},c_{\rm del}>0$ such that, whenever $m\ge C_0NJ$ and
$C_{\rm pc}\sqrt\rho\le r\le1/8$,
\cref{alg:hybrid-phase} returns, with probability at least
$1-\zeta$, a nonnegative reweighting $A'$ with unchanged row and
column sums, support contained in that of $A$, and
$|\supp A'|\le(1-c_{\rm del})m$, whose error satisfies
\[
 \|R_L(A)^{\dagger/2}(A'-A)R_R(A)^{\dagger/2}\|\le4r.
\]
The change satisfies the strong-SV kernel conditions. The running
time is $\softO(m+N)$, with factors polynomial in $\log(2n)$ and
$\log(1/\zeta)$ and no polynomial inverse-radius factor.
\end{lemma}
\begin{proof}
In the first branch, \cref{lem:fast-phase} gives all output
guarantees. Its running time satisfies
\begin{align}
 T_{\rm sparse}
 &\le\softO\bigl((m+N)r^{-27/2}\bigr)\notag\\
 &\le\softO\bigl((m+N)\rho^{-27/4}\bigr)\notag\\
 &\le\softO\bigl((m+N)\ell^{270}\bigr)
 =\softO(m+N).
 \label{eq:sparse-branch-time}
\end{align}

Consider the dense branch, conditional on successful decomposition
and implicit-factor construction. We have $q\ge m/8$,
$\sum_H|V(H)|\le NJ$, and $\phi_c=\Theta(\ell^{-3})$.
The degree lower bound of \cref{lem:dense-cores} gives
$d_0\ge c\rho^{-1}\ell^{-4}\ge(c/c_{\rm dense})\ell^{36}$, while
$\phi_c^{-4}\log^3(2m)\le C\ell^{15}$.
Choosing $c_{\rm dense}$ sufficiently small verifies the hypothesis
of \cref{lem:dense-cube}. It also makes the number of edges large
enough for its stated constant-probability saturation guarantee.

Let $x_*=P_Kg$, where $K=S\cap[-1,1]^F$. By
\cref{lem:dense-gaussian,lem:cube-comparison},
$\mathbb E\|\overline{\mathcal X}(x_*)\|\le C_G\sqrt\rho$. Markov's
inequality then gives
\[
 \Pr\!\left[\|\overline{\mathcal X}(x_*)\|>r/8\right]
 \le\frac{8C_G\sqrt\rho}{r}
 \le\frac{8C_G}{C_{\rm pc}}
 \le\frac1{20}.
\]
On the cube-projection event, which we take to have probability at
least $9/10$, the returned $y$ satisfies
$\|y-x_*\|_2\le\tau$, and a constant fraction of the coordinates
of $y$ have magnitude at least $1-\tau$. Choose $\gamma_c>0$ so
that, after the sign choice, at least $\gamma_c q$ coordinates of
$y$ lie below $-1+\beta$, using $\tau\le\beta$.

For any $h\in S$, the implicit projected-variance estimate gives
\begin{align*}
 \|\overline{\mathcal X}(h)\|
 &=\left\|\sum_eh_e\overline{\mathcal X}(\Pi e_e)\right\|\\
 &\le\|h\|_2
       \left\|\sum_e\overline{\mathcal X}(\Pi e_e)^2
                                             \right\|^{1/2}\\
 &\le\sqrt{2\rho}\,\|h\|_2.
\end{align*}
Since $y-x_*\in S$, we obtain, on the intersection of these events,
\[
 \|\overline{\mathcal X}(y)\|
 \le r/8+\sqrt{2\rho}\,\tau
 \le r/4,
\]
where the second inequality uses $\tau\le1$ and a further absolute
increase in $C_{\rm pc}$. Hence $U\le2\|\overline{\mathcal X}(y)\|\le r/2$.
Thus both checks pass on an event of probability at least
$1-1/10-1/20=17/20\ge3/4$.
For fresh Gaussian trials, conditional on the fixed pieces and
factors and on any preceding trials, this lower bound is unchanged.
Consequently,
\begin{align*}
 \Pr[\text{no trial passes}\mid\text{valid pieces and factors}]
 &\le(1/4)^R\le\zeta/4.
\end{align*}

On the event that all norm estimates satisfy their lower guarantee,
an accepted vector obeys
\begin{align*}
 \|\mathcal X(y)\|
 &\le2\|\overline{\mathcal X}(y)\|
 \le2U\le r.
\end{align*}
The cycle-rounding proof applies to these cores with their overlap
bound and envelope $\Lambda_c$. Taking $C_0\ge16/\gamma_c$ gives
\[
 m-|\supp A'|
 \ge\frac{\gamma_c q-NJ}{2}
 \ge\frac{\gamma_c q}{4}
 \ge\frac{\gamma_c m}{32},
\]
and the total SV error is at most $r+2\beta\Lambda_c=r+r/4\le4r$.
Both changes preserve each core's degrees. The cancellation in
\eqref{eq:double-center-cancel} and the projected-variance lemma
also give the kernel conditions. Use the smaller deletion constant
of the two branches to define $c_{\rm del}$.

For the dense running time, every matrix-vector product costs
$\softO(m+N)$ by \cref{lem:implicit-factors}. The covariance
matrices and projectors in the proof never need to be formed.
The capped cube routine costs
$\softO(m\log(2m/\tau))$ per trial. Since $m\le n^2$, $N\ge1$,
and $r\ge C_{\rm pc}\sqrt\rho$, we have
$\rho=16N\log(N+1)/m\ge cn^{-2}$. Hence
$\tau=r/(32\Lambda_c)\ge c/(n\ell^6)$ and $\log(2m/\tau)=O(\log(2n))$.
Each norm check uses $O(\log(2m/\delta))$ products.
Pruning, exact tree correction, and cycle rounding take
$\softO(m+N)$ time. With $R=O(\log(4/\zeta))$, their total is
$\softO(m+N)$ as claimed.

To account for failure, let $E_D,E_Z,E_R$ denote failure of the
decomposition, the factors, and all trials, respectively. Let
$E_j$ denote an invalid lower guarantee in the $j$th norm test,
if that test occurs. Conditional on its input, $\Pr[E_j]\le\delta$.
Thus
\begin{align*}
 \Pr[\text{some output guarantee fails}]
 &\le\Pr[E_D]+\Pr[E_Z\mid E_D^c]
      +\Pr[E_R\mid E_D^c\cap E_Z^c]
      +\sum_{j=1}^R\Pr[E_j]\notag\\
 &\le\zeta/4+\zeta/4+\zeta/4+R\delta
 =\zeta.
\end{align*}
The sparse branch has failure probability at most $\zeta$ by
\cref{lem:fast-phase}.
\end{proof}

\subsection{Error, support bound, and running time}

\begin{proof}[Proof of \cref{thm:fast}]
Run \cref{alg:fast-sv}. Returning $A_0$ handles the cases $m_0=0$,
$n=1$, and $m_0\le M$. If $\varepsilon<n^{-1/2}$, it also meets
the support bound because $m_0\le n^2<n/\varepsilon^2$.
We therefore assume $n\ge2$, $\varepsilon\ge n^{-1/2}$,
and $m_0>M$.

First suppose every randomized call meets its guarantee.
We bound the failure probability at the end.
As in the existence proof, we verify each grid rounding by
induction. Before a phase, all preceding roundings have placed the
positive weights in $[1/K_g,n]$, and the initial weights are one.
Thus $W_t\le nK_g$ and $J_t\le C_J\log(2n)$ for an absolute
$C_J$. Set $\vartheta=1-c_{\rm del}$ and
$\Gamma=n\log(2n)$. Choose
\[
 C_*\ge\max\left\{2C_0C_J,\,
           \left(\frac{1024C_{\rm pc}}{1-\sqrt\vartheta}\right)^2,\,
           1\right\}.
\]
Whenever $m_t>M$, this choice gives $m_t>M\ge C_0N_tJ_t$ and
$\rho_t=16N_t\log(N_t+1)/m_t\le64\Gamma/m_t\le64\varepsilon^2/C_*$.
Hence
\begin{align*}
 r_t&=C_{\rm pc}\sqrt{\rho_t}
                    +\frac{\varepsilon}{128T_{\max}}\\
 &\le\frac{8C_{\rm pc}\varepsilon}{\sqrt{C_*}}
                    +\frac{\varepsilon}{128T_{\max}}
 \le\frac{\varepsilon}{64}<\frac18.
\end{align*}
The nearly linear phase lemma, \cref{lem:hybrid-phase}, therefore
applies. Each completed phase
satisfies $m_{t+1}\le\vartheta m_t$, including its grid rounding.
Consequently $m_t\le\vartheta^t m_0$, and the algorithm runs at
most $T_{\max}$ phases.

Consider a prefix of $s$ phases, possibly stopping just before
its last grid rounding. Since
$m_t\ge\vartheta^{-(s-1-t)}m_{s-1}>
\vartheta^{-(s-1-t)}M$, we have
\begin{align*}
 \sum_{t=0}^{s-1}\sqrt{\rho_t}
 &\le8\sum_{t=0}^{s-1}\sqrt{\frac{\Gamma}{m_t}}\\
 &\le8\sqrt{\frac{\Gamma}{M}}\sum_{j=0}^{s-1}\vartheta^{j/2}
 \le\frac8{1-\sqrt\vartheta}\sqrt{\frac{\Gamma}{M}}.
\end{align*}
Let $\delta_t=4r_t$ and let $\alpha_t$ be the rounding error in
the original metric, taking $\alpha_t=0$ if that rounding has
not occurred. Our choice of $C_*$ gives
\begin{align*}
 D_s:=\sum_{t<s}\delta_t
 &=4C_{\rm pc}\sum_{t<s}\sqrt{\rho_t}
                  +\frac{s\varepsilon}{32T_{\max}}\\
 &\le\frac{32C_{\rm pc}}{1-\sqrt\vartheta}
                    \sqrt{\frac{\Gamma}{M}}+\frac{\varepsilon}{32}
 \le\frac{\varepsilon}{16},
\end{align*}
and \cref{cor:one-grid} gives
$a_s:=\sum_{t<s}\alpha_t\le\min\{s,m_0\}\varepsilon/(32m_0)
\le\varepsilon/32$.
Only completed roundings enter $a_s$. Applying
\cref{lem:sv-composition} to this prefix gives
\begin{align*}
 e^{2D_s}(D_s+a_s)
 &\le e^{\varepsilon/8}
             \left(\frac{\varepsilon}{16}+\frac{\varepsilon}{32}\right)\\
 &\le\frac{3e^{1/8}}{32}\varepsilon
 <\min\{\varepsilon,1/4\}.
\end{align*}
The $1/4$ bound verifies the next grid rounding and closes the
induction. The $\varepsilon$ bound proves the final approximation
guarantee.

The algorithm stops with at most
$M=\lceil C_*n\log(2n)/\varepsilon^2\rceil$ arcs, never increases
support, and preserves all row and column sums. Every final weight
is a multiple of $1/K_g$ in $[0,n]$, so it has
$O(\log n+\log(1/\varepsilon))$ bits.

For the running time, use \cref{lem:hybrid-phase},
$T_{\max}=O(\log(2m_0))$, and $N_t\le2m_t$. The per-phase failure
probability $\eta/T_{\max}$ contributes only logarithmic factors.
Since $\sum_t m_t\le m_0\sum_{t\ge0}\vartheta^t=m_0/(1-\vartheta)$,
\begin{align*}
 \sum_t T_{\rm phase}(m_t,N_t,r_t,\eta/T_{\max})
 &\le\sum_t\softO(m_t+N_t)\\
 &\le\softO\left(\sum_t m_t\right)
 =\softO(m_0).
\end{align*}
Each grid rounding computes a maximum spanning forest, rounds
nontree weights down, and traverses each tree once. This costs
$O(m_t\log(2n))$ per phase. Reading the input and keeping isolated
vertices adds $O(m_0+n)$, giving the claimed total time.

Finally, let $F_t$ be the event that phase $t$ is the first to
report failure or violate a subroutine guarantee. Conditional
on all preceding phases succeeding, its failure probability is
at most $\eta/T_{\max}$. Therefore
\begin{align*}
 \Pr[\text{some phase is unsuccessful}]
 &=\sum_{t<T_{\max}}\Pr[F_t]\\
 &\le\sum_{t<T_{\max}}
       \Pr[\text{all preceding phases succeed}]
                               \frac{\eta}{T_{\max}}\\
 &\le T_{\max}\frac{\eta}{T_{\max}}
 =\eta.
\end{align*}
Thus all output guarantees hold together with probability at
least $1-\eta$.
\end{proof}

\section{Conclusion and future work}

We showed that every unweighted directed graph has a strong SV sparsifier with $O(n\log(2n)/\varepsilon^2)$ reweighted original arcs and exactly preserved degrees, and that such a sparsifier can be found in nearly linear time. The main new ingredient is a projected variance bound in which each resistance piece contributes only its own Schur-complement defect. This lets the matrix discrepancy tools apply without any loss from vertices shared between pieces.

There are several avenues for future work. Perhaps the most natural is to remove the remaining logarithm. For undirected graphs, Batson, Spielman, and Srivastava~\cite{bss} achieve $O(n/\varepsilon^2)$ edges, and \cref{prop:lower} shows that $O(n/\varepsilon^2)$ arcs would also be optimal in the directed setting. In our argument the logarithm enters through the resistance parameter of the effective-resistance decomposition, so closing the gap will likely require a different decomposition, or perhaps a potential-function argument in the spirit of~\cite{bss}.

It would also be interesting to understand whether the additive $n\log(2U)$ term in \cref{thm:weighted} is really needed, and whether the weighted bound can be achieved in nearly linear time as well. On the algorithmic side, our construction is randomized, works in the real-arithmetic model, and hides large polylogarithmic factors. A deterministic algorithm with the same support bound, a word-RAM analysis, or simply a smaller polylogarithmic overhead would each make the result more practical. Finally, we hope that strong SV sparsifiers can be maintained under arc insertions and deletions, much as Zhao~\cite{zhao-dynamic} does for weaker notions of directed approximation.

\section*{AI disclosure}

We interacted with GPT-6 Astra to explore whether our expander-based idea could improve the results in this direction. Over a series of conversations, this led to a pro1
of idea, and Claude Opus~5.5 then helped the author flesh out the writing of each lemma and proof.

\section*{Acknowledgements}

EA is supported by NSF Grant CCF-2338816, and thanks Albert Weng for feedback on the manuscript, and Zhao Song for insightful discussions and for sharing his result on spectral
sparsifiers.

\bibliographystyle{alpha}
\bibliography{main}

\end{document}